\documentclass[utf8]{article}

\usepackage{arxiv}
\usepackage{natbib}

\usepackage[utf8]{inputenc} 
\usepackage[T1]{fontenc}    
\usepackage{hyperref}       
\usepackage{url}            
\usepackage{booktabs}       
\usepackage{amsfonts}       
\usepackage{nicefrac}       
\usepackage{microtype}      
\usepackage{lipsum}
\usepackage{subfiles}
\usepackage{graphicx, xcolor}
\usepackage{mparhack}
\usepackage{subcaption}
\usepackage{placeins}
\usepackage{todonotes}
\usepackage[export]{adjustbox}
\graphicspath{ {./figures/} }

\makeatletter
\@mparswitchfalse
\makeatother
\normalmarginpar

\usepackage{soul}
\usepackage{color, colortbl}

\usepackage{amsmath}
\numberwithin{equation}{section}

    \usepackage{etoolbox,refcount}
    \usepackage{multicol}
    \usepackage{scalerel,amssymb}
    
    \newcounter{countitems}
    \newcounter{nextitemizecount}
    \newcommand{\setupcountitems}{%
      \stepcounter{nextitemizecount}%
      \setcounter{countitems}{0}%
      \preto\item{\stepcounter{countitems}}%
    }
    \makeatletter
    \newcommand{\computecountitems}{%
      \edef\@currentlabel{\number\c@countitems}%
      \label{countitems@\number\numexpr\value{nextitemizecount}-1\relax}%
    }
    \newcommand{\nextitemizecount}{%
      \getrefnumber{countitems@\number\c@nextitemizecount}%
    }
    \newcommand{\previtemizecount}{%
      \getrefnumber{countitems@\number\numexpr\value{nextitemizecount}-1\relax}%
    }
    \makeatother    
    \newenvironment{AutoMultiColItemize}{%
    \ifnumcomp{\nextitemizecount}{>}{1}{\begin{multicols}{3}}{}%
    \setupcountitems\begin{itemize}}%
    {\end{itemize}%
    \unskip\computecountitems\ifnumcomp{\previtemizecount}{>}{1}{\end{multicols}}{}}

\usepackage{multirow}
\usepackage{pdflscape}
\usepackage{float}

\usepackage{cleveref}  

\usepackage{subcaption}

\definecolor{highlightTable}{gray}{0.9}

\newcommand{\set}[1]{\{#1\}}

\newcommand{\revisionEnhancement}[1]{#1}

\newcommand{\makeAuthor}[4]{#1\\#2\\#3\\\texttt{#4}\\}
\newcommand{\makeKeywords}{
virtual reality \and 
haptic feedback \and
causal learning \and
causal discovery \and
causal representation \and
sensorimotor behavior \and
dual system theory
}

\title{Cause-effect perception in an object place task}

 \author{
  \makeAuthor{Nikolai Bahr}{Cognitive Neuroinformatics, University of Bremen}{Bremen, Germany}{nibahr@uni-bremen.de}
  \And
  \makeAuthor{Christoph Zetzsche}{Cognitive Neuroinformatics, University of Bremen}{Bremen, Germany}{\relax}
  \And
  \makeAuthor{Jaime Maldonado}{Cognitive Neuroinformatics, University of Bremen}{Bremen, Germany}{\relax}
  \And
  \makeAuthor{Kerstin schill}{Cognitive Neuroinformatics, University of Bremen}{Bremen, Germany}{\relax}
 }

\begin{document}
\maketitle


\begin{abstract}
\revisionEnhancement{
In this paper, we conducted an exploratory study in virtual reality to investigate whether people can discover causal relations in a realistic sensorimotor context and how such learning is represented at different processing levels (conscious‑cognitive vs. sensorimotor). Additionally, we explored the relationship between human causal learning and state-of-the-art causal discovery algorithms. The task consisted of placing a glass on a surface. To enhance ecological validity, the setup included haptic rendering to simulate the glass's weight and contact force. The glass would break if the contact force exceeded its breakability threshold, determined by the causal structure $weight \rightarrow breakability \leftarrow color$. Participants were asked to repeatedly transport and place glasses of varying weights and colors on a surface without breaking them. Therefore, to accomplish the task, participants had to discover the underlying causal structure. The trials were conducted over three separate sessions, each aimed to capture a different behavior ((i) naive and causally unaware, (ii) exploratory, and (iii) consolidated and causally aware). After each session, participants completed a questionnaire providing a measure of their conscious understanding of the task's causal structure. Sensorimotor representations were inferred by applying three causal‑discovery algorithms (PC, FCI, FGES) to the recorded trial‑by‑trial variables, and conditional mutual information was used to quantify the strength of causal influence on the sensorimotor level.
Results show that (i) participants identified the weight‑breakability link ($\approx 76\%$ correct after the final session) and, to a lesser extent, the color‑breakability link ($\approx 43 \%$ correct), but they could not reliably infer causal direction. (ii) Sensorimotor analysis revealed a robust weight‑force coupling that increased across sessions, whereas the color‑force coupling was weak and noisy, yet mutual information indicated an attempted learning. (iii) Discovery algorithms recovered the underlying structure across the three sessions. Together, these findings indicate that humans can, to some extent, perceive the causal structure of the task and that conscious and sensorimotor representations are partially dissociated.
} 
\end{abstract}


\section{Introduction}
\revisionEnhancement{
 Acting intelligently in the world in everyday life is impossible without the appropriate perception and understanding of the causal structure of the physical environment.
 (Is the food spoiled because of the sunlight or the heat of the room? 
 What happens if I plug in the cable of the coffee machine?
 If my knife gets stuck while slicing bread, should I strengthen the pressure or increase my movement? Is the pavement wet because of the sprinkler or the rain?) 
 The inference of causal relations provides the basis for reasonable decisions and a means to adapt to changes and contingencies in the environment and task conditions~\citep{hattori2007covariationDetection, rottman2014reasoning}.
 Causal understanding enables the inference of the effect of actions taken and provides the basis for controlling movements~\citep{glymour2003causalKnowledgeBayes,shams2022bayesiancausalInference}. 
 Cause-effect relationships tend to interact in a sparse/local way and are mostly invariant under domain shifts, thereby improving the stability of decision-making systems that use  causal knowledge instead of statistical associations as a basis.
Understanding the causal structure of the world is a key challenge for both humans and artificial systems, like robots or machine learning algorithms~\citep{Hellstroem2021causationRobotics,Scholkopf2021representationLearning}.

A crucial first step for a successful causality-based behavior is the \emph{identification} of the causal structure of a system, also known as causal structure learning, or causal discovery \citep{glymour2019discovery,nogueira2022discovery}. 
For this identification, one must first disentangle spurious associations from actual cause-effect relationships. Suppose you clap your hands and, at the same moment, hear a dog bark outside. To figure out whether your clapping caused the dog to bark or if the two events simply happened at the same time by chance, you need to separate a real cause-effect link from a spurious association. Without further evidence, it would be mistaken to assume your clapping was responsible for the barking, rather than recognizing that the two events may have occurred independently. Furthermore, one has to take probabilistic dependencies into account, since only a smaller part of the relevant causal relations can be reduced to deterministic rules.


There are numerous algorithms to automate the identification, which are formally sound and provably converge towards the optimal solution. 
However, regardless of these desirable properties, there are still no convincing approaches to the autonomous handling of everyday tasks. 
Part of this deficit in robotic applications may originate from necessary assumptions that are often not strictly met in practice, e.g., the representation of the causal system must be provided (relevant variables, etc.) or assumptions about the functional form of the causal mechanisms. Additionally, working with large sets of variables is particularly challenging due to the exponential growth in combinatorial space.
}

\revisionEnhancement{In contrast to robots, though, humans possess full everyday competence in spite of facing the same problems.
It is thus of interest to compare the properties of the causal processing of humans and of machine algorithms.
In psychological research on causality the experimental setups are often relatively abstract, involving for example a hypothetical communication between aliens ~\citep{steyvers2003inferring,rottman2014reasoning}, the operation of machines like the "blicket detector" \citealp{Lucas2010formHierarchicalBayesian,goddu2024developmentHumanCausal}, the hypothetical influence of temperature and pressure on a rocket launch  \citep{lagnado2002learning},  or the characteristics of fictitious types of stars~\citep{rehder2017induction}. 
}

\revisionEnhancement{In this work, we want to investigate how humans can learn causal relationships on a more basic level in the context of a realistic sensorimotor task.
In particular, we want to investigate the following set of interrelated research questions: 

\begin{itemize}
    \item Can causal relations in a natural sensorimotor task of object transport be perceived and identified by human subjects?

   \item Does this causal learning take place on different levels of abstraction and processing?
    
    \item How can we measure the human causal representation in this context?
 
    \item How does the causal representation evolve over the course of an experiment?

    \item Can we compare the human learning in such a sensorimotor task with the operation of machine algorithms?
\end{itemize}

For this, we develop a suitable experimental setup in this work which (i) feels natural to our subjects, in contrast to the sometimes artificial setups used in other human causal learning experiments, (ii) is firmly embedded in a behaviorally meaningful sensorimotor task, as opposed to the abstract reasoning problems being often used, (iii) allows us to implement a ground truth causal structure of our free choice, and (iv) provides us with measurements of human motor data which allow us to analyze the underlying causal sensorimotor representation.
}

\section{Background and Related Work}\label{sec:related_work}
\revisionEnhancement{
\subsection{Causal Bayesian Networks and Causal Discovery}\label{sec:bayesian_networks_related_work}

In this work, we use the formalism of causal Bayesian networks~\citep{pearl2009bayesianNetworks}. A causal Bayesian network is a probabilistic graphical model that represents the cause-and-effect relationships among a set of variables using a directed acyclic graph (DAG), where nodes represent the variables and directed edges direct causal relations, visually represented as arrows. The nodes can represent categorical, discrete, and/or continuous variables. An arrow represents the statistical dependency between variables. 

The notion of the strength of a causal effect refers to the degree to which changing one variable through manipulation influences another variable. A strong causal effect indicates that a manipulation yields substantial variation in the outcome, while a weak effect indicates minimal impact on the outcome. Using the do notation, the general causal effect is defined as ${P(Y = y|do(X = x))}$, which denotes the conditional probability of $Y = y$, given the intervention $do(X = x)$~\citep[Sec. 3.2]{pearl2016causalStatistics}.
\revisionEnhancement{An intervention describes the active manipulation of the given system, i.e., \emph{setting} a variable $X$ to the value $x$ independently of its causes instead of \emph{selecting} data in which $X = x$ (and therefore dependent on its causes).}
\cite{janzing2013mutualinformation} have reviewed other measures of causal strength that have been proposed in the literature, emphasizing that the applicability of a particular measure depends on assumptions of the functional dependency between variables and the structure of the DAG (we will return to this topic in Section \ref{subsec:methods-causal-strength-measure}, where we discuss the measure of causal strength used in our setup).

The quantification of causal effects relies on the availability of a DAG. When the DAG is not known, it can be identified from data using causal discovery algorithms (also known as structure learning algorithms) (for recent accounts on the subject, see the reviews by \cite{glymour2019discovery} and  \cite{nogueira2022discovery}. Causal discovery algorithms systematically analyze many possible causal structures in relation to the statistical properties of the data. The causal structure can be identified by using conditional independence tests or model scoring~\cite{glymour2019discovery}. For example, in the case of three variables, the common-effect structure (e.g., $C_1 \rightarrow E \leftarrow C_2$, the one we use in this paper) has a unique set of independence statements~\citep{pearl2009discovery}.}

There are two well-established approaches to causal discovery: constraint-based and score-based algorithms. Constraint-based algorithms identify the causal structure by testing for probabilistic (in)dependencies between the variables \citep{glymour2019discovery}. Since this is done repeatedly, the process models a multiple-testing problem as opposed to score-based approaches~\citep{spirtes2010introduction}. However, most score-based algorithms require stronger assumptions over the data-generating process than constraint-based algorithms~\citep{huang2018generalized}.
Score-based algorithms identify the causal structure by optimizing a score function such as the Bayesian Information Criterion (BIC), which approximates the posterior probability of the causal structure given the data~\citep{malinsky2017discovery,glymour2019discovery}.
\revisionEnhancement{A special class of score-based algorithms is Bayesian causal discovery~\citep[BCD,][]{heckerman2006bayesian}. The algorithms provide a degree of belief of every hypothetically possible causal graph for the given setting in the form of the Bayesian posterior. An advantage of BCD is that incorporating prior knowledge is easily possible. However, there are also risks associated with a proper choice of the prior~\citep{weakliem1999critique}. Compared to constraint-based approaches, the complexity of BCD is high, and finding the highest scoring DAG model is generally NP-hard~\citep{chickering1996learning}.
In particular, it is impossible to apply standard BCD to configurations with more than six variables~\cite{karimi2024challengesBCD} 
To mitigate this, approximations using Monte-Carlo methods or Gaussian approximations can be done in BCD~\citep{heckerman2006bayesian}. Additionally, score-based methods such as the GES algorithm employ search heuristics or greedy schemes~\citep{spirtes2001causation}.
}
\revisionEnhancement{
\subsection{Human Causal Inference}\label{sec:human_causal_inference_related_work}
It has been suggested that research on human causal learning and reasoning should make use of the formal description in terms of causal Bayesian networks (e.g.,  \citep{waldmann1998bayesian,glymour2001mindArrow,glymour2003causalKnowledgeBayes,gopnik2004causalLearningChildren,tenenbaum2006theorybasedBayesian}.
Meanwhile, a majority of studies of human causal inference make  use of causal Bayesian networks as the normative framework, and a number of researchers applied it successfully as a computational framework to study how humans infer causality by analyzing patterns of correlation and the impacts of interventions (e.g.,  \citep{lagnado2002learning,glymour2003causalKnowledgeBayes,steyvers2003inferring,gopnik2007bayesianLearningCognitiveDevelopment,meder2010observingAndIntervening,rottman2014reasoning}).

\subsubsection{Causal inference in cognitive reasoning}

Most studies of human causal inference focus on the cognitive level of analysis. This aims to investigate human qualitative and quantitative causal judgments and provides a basis to investigate whether people follow
the causal Markov assumption and whether people make use of the parameters of the causal structures~\citep{glymour2003causalKnowledgeBayes, rottman2014reasoning}, as well as to investigate the effect of causal knowledge on categorization (i.e., process of inferring an object's category membership given its features)~\citep{rehder2017categorization} and category-based induction (i.e., the process of inferring features given an object's category membership)~\citep{rehder2017induction}.

A key problem in experiments of causal learning is the unpredictable influence of previous experience.
It has been noted that, in experimental settings, it is difficult to assess or control for potential contributions to judgment or behavior coming from participants' prior real-world knowledge, experience, or beliefs~\citep{rottman2014reasoning,bailey2024causalInferenceOnHumanBehavior}. 
A common approach in cognitive causal research thus consists of providing participants with a cover story describing a specific context of the variables involved and the inference tasks. Typically, unnaturalistic tasks ( e.g., the blickets paradigm where participants can intervene on objects that light up or play music \citealp{Lucas2010formHierarchicalBayesian,goddu2024developmentHumanCausal}), unfamiliar contexts (e.g., rocket launch \citep{lagnado2002learning}), or artificial contexts are provided (e.g., fictitious types of stars~\citep{rehder2017induction} or telepathic aliens who transfer their thoughts through mind reading~\citep{steyvers2003inferring,rottman2014reasoning}).

Complete learning of a causal representation requires knowledge about its structure (the directed graph) and its parameters (functional relationships).
Early causal research in psychology has concentrated on the latter, for example, by describing the perceived strength of a causal influence using the $\Delta P$ rule~\citep{Jenkins1965,Allan1980,Cheng1990}, which is a measure of the strength of covariation between a (cause) variable and a subsequent (effect) variable,  and the powerPC theory~\citep{Cheng1997}. The $\Delta P$ rule has been used as a normative measure to assess human judgements of causal association between variables (see \citep[Chap. 4]{glymour2001mindArrow},  \citep{hattori2007covariationDetection}, and references therein).

Regarding the structure of the causal graph, research has been focused on paradigmatic causal structures such as chains, common effects, and common causes~\citep{rottman2014reasoning,rehder2017categorization}; typically involving binary causes and effects~\citep{glymour2001mindArrow,rottman2014reasoning}. 

Judgments often align with causal direction, yet subjects show limited sensitivity to causal parameters and structure~\citep{rottman2014reasoning}. When informed about causal strength, their predictions fall short of normative expectations. This insensitivity manifests as violations of the Markov condition, where conditionally independent variables are viewed as causally dependent~\citep{rottman2014reasoning,rehder2017induction}. Proposed explanations include assumptions of hidden causal mechanisms and the influence of prior knowledge or beliefs that introduce unconsidered causal links~\citep{rottman2014reasoning,rehder2017induction}.

Humans, despite their imperfections, are surprisingly quick at learning causal relations, faster than expected had they used full conditional probability distributions. According to \cite{hattori2007covariationDetection}, this process begins with extracting covariation information from observations, which helps eliminate unrelated events as potential causes. They argue that after detecting covariation, an analytic process involving domain-specific causal knowledge and interventions follows, aiding in distinguishing real from spurious causes. \cite{steyvers2003inferring} demonstrated that interventions outperform observation-only methods for identifying causal structures in humans, while \citep{lagnado2004advantage} emphasize the importance of temporal cues in interpreting intervention results. The surprisingly fast learning, and the ability to incorporate the aforementioned influence of prior experience have been used as arguments for considering Bayesian causal discovery for the modeling of human causal learning \citealp{waldmann1998bayesian,griffiths2009theoryBasedCausalInduction}, where domain-specific prior knowledge guides statistical inference (termed \textit{theory-based causal induction}) \citealp{tenenbaum2006theorybasedBayesian,griffiths2009theoryBasedCausalInduction}.
}

\revisionEnhancement{
\subsubsection{Causal inference in perceptual reasoning}
The Bayes' rule and Bayesian networks formalisms have been used as normative models to investigate human performance in various perceptual and sensorimotor tasks \citep{shams2010causalPerception,shams2022bayesiancausalInference}. In particular, Bayesian formalisms are used to investigate two problems in multisensory perception \citep{shams2022bayesiancausalInference}: determining the causal structure that generated the sensory cues among a set of mutually exclusive options (termed the \textit{Bayesian Causal Inference model}) and estimating hidden variables given sensory cues (termed the \textit{integration problem}).  

The Bayesian Causal Inference model is a statistical framework that determines the more likely causal structure between two options (commonly a common cause vs. independent causes) \citep{shams2022bayesiancausalInference}. A key example is the research by \cite{koerding2007causalMultisensory}, which explored how subjects infer common or distinct causes for auditory and visual cues. A common cause (e.g., $X_1 \leftarrow S \rightarrow X_2$) is inferred when auditory and visual signals are perceived as close, while distinct causes (e.g., $S_1 \rightarrow X_1$ and $S_2 \rightarrow X_2$) are identified when they are perceived as distant. These causal structures are viewed as \textit{competitive priors} \citep{shams2022bayesiancausalInference}, where prior beliefs influence causal interpretation, and these priors can be shaped by experience or evolution \citep{shams2022bayesiancausalInference}. Experimental data suggest that the Bayesian Causal Inference model effectively captures human behavior across various tasks (see \cite{shams2022bayesiancausalInference}). However, it's uncertain whether this inference happens consciously or if the nervous system actively commits to a causal structure during processing \citep{shams2022bayesiancausalInference}. Moreover, the competitive prior structures are assumed to be fixed, making the inference a matter of choosing one structure from a limited set of options.

Regarding the general causality literature, the term causal inference relates to computations based on a known causal structure, as data alone cannot resolve causal questions without understanding the data-generating process \citep{pearl2016causalStatistics}. Causal discovery, or causal structure learning, involves identifying the causal structure of a process from its statistical properties without assuming any specific structures \textit{a priori} \citep{glymour2019discovery,nogueira2022discovery}. Thus, the Bayesian Causal Inference model of human perception, which uses competitive prior structures, differs from general causal discovery/learning, although the conceptual resemblance is acknowledged (see \cref{sec:bci_discussion} in the appendix). The term \textit{Bayesian probabilistic inference} describes how humans assess belief over hypothetical causal structures based on data \citep{gopnik2007bayesianLearningCognitiveDevelopment}.
}
\section{Methods}\label{sec:methods}

\revisionEnhancement{
\subsection{Motivation}\label{subsec:methods-motivation}

In the introduction, we have outlined the main questions we aim to address through our investigation.
In this section, we aim to outline the various aspects of our motivation and examine the detailed requirements that must be met to facilitate a thorough investigation.
Here is an overview of the problems and requirements we will discuss in this and the following sections:
}

\revisionEnhancement{
\begin{enumerate}

    \item \label{item:observation-only_identification} \emph{Observation-only identification.} To avoid problems with the role of interventions, we seek an experimental setup with a causal structure that can be uniquely identified from observations only.

    \item \label{item:levels_causal_representation} \emph{Different levels of causal representation.} Is there one unique causal representation? Or do we have reason to believe that there exist different levels of causal representations, in particular for the handling of a sensorimotor task?
    
    \item \label{item:perceived_causal_representation} \emph{Perceived cognitive causal representation.} We have to find means to determine which causal representation our subjects use on the conscious cognitive level.

    \item \label{item:sensorimotor_causal_representation} \emph{Sensorimotor causal representation.} Is it different from the causal representation on the perceptual level? How can we measure this representation?

    \item \label{item:sensorimotor_relevance} \emph{Sensorimotor relevance.} How can we ensure that the causal structure is relevant for the sensorimotor level?

    \item \label{item:observable_sensorimotor_representation} \emph{Observability of the sensorimotor causal representation.}  The internal variables of representations can usually not be observed. 
    
    \item \label{item:problems_naturalness} \emph{Problems of naturalness.} Natural tasks pose a problem for causal experiments due to the possible influence of long-term experience.

    \item \label{item:learning effect} \emph{Measure the learning effect.} How can we measure how much subjects learn about the causal relations, both on the conscious and on the sensorimotor level? How can we deal with prior experience?
    
    \item \label{item:machine_comparison} \emph{Comparison of human causal learning with machine algorithms.} Do we need access to a full-fledged robot? How can we enable a fair comparison?   

\end{enumerate}
}

\begin{figure*}[htbp]
\centerline{\includegraphics[clip, width=0.5\textwidth]{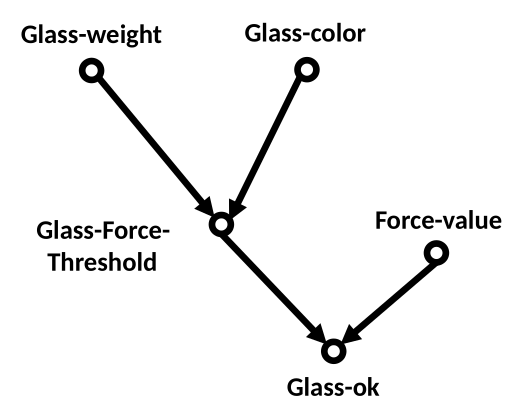}}
\caption{\revisionEnhancement{The ground truth causal structure of the experimental setup. Upper left sub-graph: causal dependencies between glass properties. Lower subgraph: causal determinants of possible glass destruction.}}
\label{figure:ground_truth}
\end{figure*}

\revisionEnhancement{
In the following, we provide detailed comments on how we considered the above-stated problems in designing the experiment and in the methods used for data analysis.
For this, we developed an experimental setup in Virtual Reality (VR) in which human subjects perform the natural task of moving an object and placing it at a prescribed target location.
In particular, the subjects had to move a glass as fast as possible while avoiding breaking it in the final contact with the supporting surface.
Special care was taken to ensure the realism of the setup by providing high-quality haptic feedback to the subjects using a \emph{PHANToM} haptic device.

\Cref{figure:ground_truth} shows the ground truth causal structure of the experimental setup.
The upper left subgraph describes the causal dependencies between the glass properties (weight, color, and force threshold). Throughout the text, we will use the terms \textit{force threshold}, \textit{breakability}, and \textit{breakability threshold} interchangeably.   
Whether these dependencies can  be recognized by our subjects is the central question of our study.
The lower part of the graph describes how the combined influence of the causes breakability and force-value determines whether the glass will remain intact (i.e., glass-ok=true) or will be broken (i.e., glass-ok=false) by the contact force against the target plate.
}

\revisionEnhancement{
In the following, we will describe our decisions and motivations in attempting to solve the aforementioned problems.

Problem~\ref{item:observation-only_identification}: \emph{Unique identification  of the causal structure by observations only.}
In general, this is not possible for most causal structures.
However, in this experiment, the causal relations should be uniquely identifiable by observation alone, without the need to intervene in the system.
This is because neither did we want to give our subjects direct access to variables that encode properties of the VR environment, nor is it easy to implement an ideal intervention in the sense of Pearl's \emph{do} operator in motion-related parameters of the hand motion, such as velocity or force.

Therefore, we used the only basic causal configuration for which a unique identification based on observations alone is possible, the \emph{common-effect} configuration (cf., e.g., \cite{spirtes2001causation}).
Such a V-like configuration is shown in the upper left part of \Cref{figure:ground_truth}.
In particular, we used a causal dependence of the breakability of the glass on two object features: the glass weight and the glass color.
(Note that the lower part of the figure is also a V-configuration with a common effect.)

How can our subjects learn the ground truth from observation?
The first two variables, weight and color, are sensory observables for our subjects via the \emph{PHANToM} force rendering mechanism and the VR screen rendering.
The breakability can be indirectly observed through the binary outcome: whether the glass is broken or not.
}

\revisionEnhancement{
Problem~\ref{item:levels_causal_representation}: \emph{Mutiple levels of causal representation.}
Does there exist only one level on which a causal representation exists and can be learned?
In general, there is overwhelming evidence that the learning and execution of a task involves different levels of representation.

The classical view of sensorimotor learning (e.g., \cite{fitts1967}) is that it occurs as a systematic and gradual transition between two representational levels. 
In the first phase, learning is determined by cognitive control and deliberate action.
In this phase, and at this level, the learner can utilize verbal instructions and acquire declarative knowledge about the task and the appropriate strategies to cope with it.
With ongoing practice (and in a continuous fashion), more and more of the control is transferred to the motor level.
Finally, the complete motor control is performed by a fully automated procedural routine, without the need for cognitive control.
Typically, in the fully automatized state, the actor is barely aware of his automatic (re-)actions.
Tasks that initially require cognitive control can thus become automatic, freeing up cognitive resources for other tasks (cf. also the two-process theory of attention by \cite{Shiffrin1977}).

Regarding these levels of sensorimotor learning, there is an analogy to Kahneman's dual-systems theory \citep{Kahneman2012}.
Key properties of Kahneman's system 1 are: fast, intuitive, automatic, and unconscious. 
System 2 properties are: slow, deliberate, analytical, and consciously aware.
However, the dual-systems theory of Kahneman primarily addresses modes of thinking rather than levels of motor control, and the relationships between the two concepts have yet to be fully worked out in detail.
(Nevertheless, something like automatized behavior in car driving is often used as an example for a Kahneman system 1 operation.)

A dual-processing approach has also been discussed in sensorimotor learning in the form of a fast and a slow process \citep{Smith2006}, or as a distinction between explicit and implicit processes \citep{Taylor2014,McDougle2015}.
The explicit processing operates with conscious awareness, requires deliberative effort, and can be influenced by verbal instructions and declarative knowledge.
Implicit processing is closely linked to sensory input and motor output, operating on an unconscious level.
It may thus be regarded as \emph{the} sensorimotor system, as opposed to the consciously operating explicit learning and cognitive reasoning.
Of special interest for the context of this paper is that these explicit and implicit processes, while often interacting with each other, can also contribute \emph{independently} to overall sensorimotor performance \citep{Suelzenbrueck2009,DAmario2024}.
}

\revisionEnhancement{
And after all, there is also evidence from causality research itself, which suggests the existence of different levels of causal representations (although this is usually not explicitly addressed).

To recognize this, let us consider two major research paradigms that are usually employed in causality research:
On the one hand, presenting subjects with certain (often artificial and abstract) data, e.g., from a fictitious laboratory experiment or the communication patterns of aliens, and asking them purposefully directed questions about the underlying causal relations.
In terms above, these experiments can be assumed to address the conscious cognitive level of causal representation.
On the other hand, there is a line of research where causal relations are investigated that underlay certain multisensory effects (the aforementioned \emph{Bayesian Causal Inference model} \citep{koerding2007causalMultisensory}.
It is a characteristic of this multisensory processing that it is fully automatized and therefore completely "shielded" from the cognitive level ("cognitively impenetrable" \citep{Pylyshyn1999}).
For example, the position attribution in the ventriloquist effect occurs despite our cognitive knowledge that puppets do not speak, and in full awareness of the principles behind the trick.
This is clearly an example of a causal representation at a level different from the conscious cognitive level.
}

\revisionEnhancement{
In conclusion, there is substantial evidence for the existence of different processing levels, both with respect to higher cognition~\citep{Evans2008}, and to sensorimotor learning~\citep{Smith2006,Taylor2014}.
To what extent these levels overlap, reduce to two major subsystems, or represent a system-wide duality being realized in multiple subsystems is currently unclear, and we do not want to take a specific stance on this matter.
For the context of our investigation, the only important point is that we can distinguish at least \emph{two} levels which are of relevance.
On the one hand, the conscious, explicit level.
This is the level on which our subjects consciously perceive and reason about the causal dependencies in our experimental setup.
On the other hand, there is the sensorimotor level.
This is the level at which the motor parameters for the hand movement are determined in dependence on the sensory input.
}

\revisionEnhancement{
Taken together, we derive the following conclusion from the above evidence:
It is possible that there exists only one unique causal representation, with the respective information being transferred during the course of learning from an explicit and conscious level to an unconscious sensorimotor level. 
It is equally possible, though, that the subsystems have a certain independence of each other.
And even if the sensorimotor level should not be perfectly cognitively impenetrable (i.e., there would be no influence from the explicit conscious level), these levels are sufficiently well separated from each other that the possibility exists that two \emph{different} causal representations develop and exist on these levels.

This poses two questions.
First, how can we measure these different causal representations (Problems~\ref{item:perceived_causal_representation} and \ref{item:sensorimotor_causal_representation}).
And second, how can we ensure a proper and fair comparison of the representations? (Problem~\ref{item:sensorimotor_relevance}).
We start with the measurement:
}

\revisionEnhancement{
Problem~\ref{item:perceived_causal_representation}: \emph{Experimental identification of the causal representation on the higher, cognitive level.}
This is a problem that can be solved relatively easily.
We can simply \emph{ask} our subjects about their representation.
We have done this in the form of a questionnaire that participants were required to answer after each of the three experiment sessions.
Furthermore, we let them draw causal graphs.
Such an identification, achieved by asking subjects about their perceived causal relations, is a standard procedure in human causality research~\citep{glymour2001mindArrow,rottman2014reasoning}.
A detailed description of the questions can be found in section~\ref{sect:questionnaire}.
}

\revisionEnhancement{
Problem~\ref{item:sensorimotor_causal_representation}: \emph{Identification of the sensorimotor causal representation.}
This is a more tricky issue since the representation cannot be directly inferred by questioning our subjects or by performing a test after each experiment session.
Of course, we have asked our subjects about the criteria they used to determine their motor strategy.
However, by the very definition of the sensorimotor level, we cannot expect our subjects to have complete conscious access to the underlying causal representation.
In particular, there is a high probability that they would confuse their sensorimotor representation with their conscious causal representation.
Therefore, we have to look for an alternative, "neutral" measuring method.
In this paper, we suggest using \emph{causal discovery methods} for identifying the sensorimotor causal representation.
To our knowledge, this is the first time a causal representation learned by humans has been analyzed using causal discovery algorithms.
(An important precondition for such an application, the availability of sufficient statistical data from our subjects, is discussed below.)
}

\begin{figure*}[htbp]
\centerline{\includegraphics[clip, width=\textwidth]{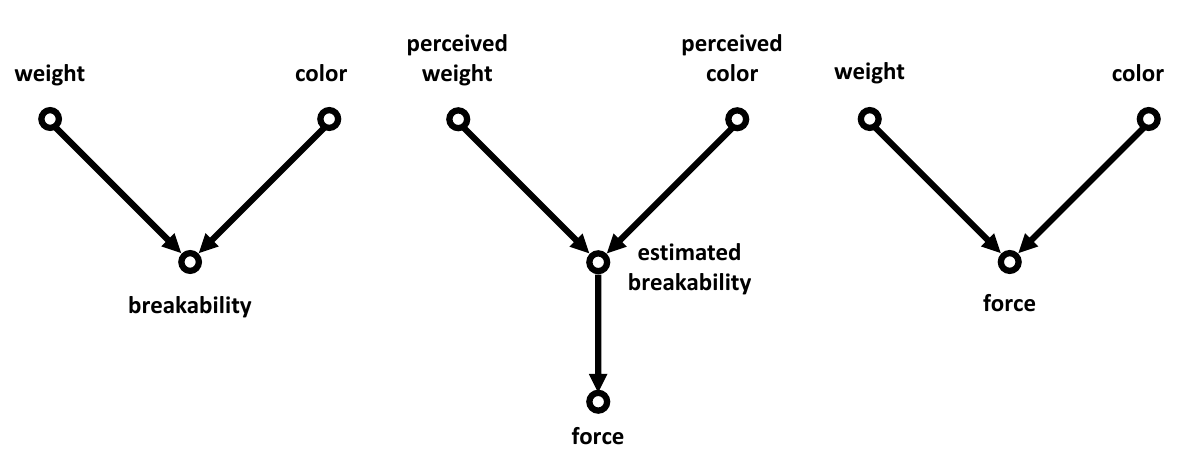}}
\caption{\revisionEnhancement{Causal graphs relevant for the sensorimotor causal representation. Left: ground truth causal structure of the experimental setup. Center: Internal representation enforced to be used for sensorimotor control. Right: The corresponding simplified proxy graph of the sensorimotor causal representation.}}
\label{figure:ground_truth_vs_representation}
\end{figure*}

\revisionEnhancement{
\emph{Problem~\ref{item:sensorimotor_relevance}: Fair comparison of the representations.}
Now that we have the methods for measuring causal representations on the two different levels available, we must turn to the second of the above-mentioned problems.
Note that a fair comparison would not easily be possible if the learning of the two representations did not employ the \emph{same} information about the causal relations in the experimental setup.
How can we determine that both levels are using the same information (in our case, this should be the ground truth causal v-structure shown in the upper left part of \Cref{figure:ground_truth})?

We designed our experiment in such a way that this is achieved by the specific structure of the sensorimotor task.
A successful accomplishment of the task, being fast and breaking not too many glasses, is only possible if the underlying sensorimotor representation does properly reflect the ground truth causal structure (including its quantitative  relations, i.e., the parameters of the corresponding structural equation model).
This is shown in \Cref{figure:ground_truth_vs_representation}.
}

\revisionEnhancement{
The left graph of the figure shows the relevant ground truth, and the center graph shows the corresponding causal graph of the sensorimotor representation.
underlying causal structure of the assumed sensorimotor control as enforced by the task.
It is important to understand how the experimental design enables us to ensure that the causal representation and the sensorimotor task are aligned in such a way that the correct sensorimotor causal representation is required for an optimal solution of the sensorimotor task:

The perceived weight and perceived color are used to compute an internal estimate of the glass's breakability.
(Note that this computation may be suboptimal or may underestimate the influence of one of the factors, possibly even up to the point that one factor or both are ignored.)
The estimated breakability is then used to compute a force that is just a small margin below the breakability threshold.
This is enforced by the task, as it is only then possible to move the glass as fast as possible (high velocity means high force at the final placement) while still avoiding the glass from breaking. 
}

\revisionEnhancement{
\emph{Problem~\ref{item:observable_sensorimotor_representation}: Observability of the variables of the sensorimotor representation.}
The application of causal discovery methods for measuring sensorimotor causal representation would ideally require access to the variables.
Since these variables are \emph{internal} variables, possibly in the form of activities of dedicated neurons, we currently have no means of direct access.
However, we can manage to obtain at least an approximate or indirect access to these variables.
We solved this by finding a setup in which the variables of the internal causal representation are either sensory variables (which can be approximated by their corresponding input, which is observable) or have an observable proxy variable, as explained in more detail below.
}

\revisionEnhancement{
This specific configuration allows us to apply some approximations:
First, we can equate perceived weight and color with the true values.
This implies assuming a sufficiently high signal-to-noise ratio, thereby enabling us to ignore sensory noise.
Note that this is standard procedure in investigations of human causal representations where internal and external variables are routinely equated.
Due to this, the two variables become observables, as we can obtain their respective values directly from the experimental software.

The second approximation is that we consider the force to be a close proxy to the estimated breakability.
This is justified by the sensorimotor task, which requires the two to differ only by a small margin (see above).
Since the force value is registered by the \emph{PHANToM}, the effect variable is now also observable.

Taken together, we now have all three variables in an observable state such that we can use them as input to a causal discovery algorithm, which infers the underlying sensorimotor causal representation.
}

\revisionEnhancement{
\emph{Problem~\ref{item:problems_naturalness} Naturalness:} 
A further motivation for using these two features is that we can thus balance possible biases due to prior experience while ensuring ecological validity.
We use two realistic object properties. 
One property, the causal relation between the weight of the glass and its breakability, is compatible with prior everyday experience.
This is because the thickness of a glass is often an indicator of greater robustness, given other properties like form and geometry.
At the same time, a greater thickness implies a higher weight of the glass.
Thus, it is well possible, though not necessarily true, that at least some of our subjects can have prior experience with a causal relation between glass weight and glass breakability.

For the other property, the color of the glass, it is reasonable to assume that there is no a priori systematic relation to the breakability.
That is because glasses of every form can be produced in every color.
We thus assume that many, if not all, of our subjects will have no a priori inclination to associate the color of a glass with its breakability.
By combining both types of relations, we expect to be able to examine how prior inclination influences the identification of causal structures.
}

\revisionEnhancement{
\emph{Problem~\ref{item:learning effect} Measuring the learning effect:}
This requires the solution of several sub-problems:
Ideally, we would like to know how much prior experience subjects have regarding the causal relations in our experiment.
On the other hand, we do not want our subjects to develop biased expectations regarding our experiment.
For example, if we were to ask our subjects before the start of the experiment whether they expect the breakability of a glass to depend on its color, they could take this as a motive for paying far more attention than they would normally do to this specific relation.
On the other hand, if they have not previously experienced such a causal relation, they could interpret our question as a manipulative means to drive them towards possibly unreasonable behavior. 
They will thus refrain from declaring such a relation, even if their experience would suggest the presence of this relation in the experimental setup.  
Both types of bias would drive subjects away from the normal, everyday behavior that we want to maintain in our experiment.
}

\revisionEnhancement{
On the other hand, if we refrain from asking questions beforehand, how do we obtain the baseline against which to measure the learning in our experiment?
This baseline problem is further exacerbated by the requirements for measuring the sensorimotor representation.
Since this measurement requires experimental data, its baseline can only be obtained by actually running the experiment, and not beforehand.

A further problem with learning is posed by the ultimate goal of the sensorimotor task: move as fast as possible \emph{without} breaking glasses.
This is in conflict with learning because most information about the underlying causal relations can be obtained with a balance between breakage and its avoidance.

As a solution to these conflicting requirements, we designed a three-session experiment:
}

\revisionEnhancement{
\emph{Session I}: We deliberately abstained from any pre-experiment tests to avoid the aforementioned bias problems.
The first session has two objectives.
First, to familiarize subjects with the task and setup.
Additionally, this establishes the baseline for later comparisons.
In this session, subjects pursue the sensorimotor task but are unaware of the main goal of the study, which is to identify the underlying causal relations.
This comes only afterwards, when subjects see the questionnaire for the first time.

We are aware that setting the baseline using the experimental data obtained during this first session and the subsequent application of the questionnaire is suboptimal, as it does not accurately represent the true initial states of the subjects.
This is because a certain degree of learning and some conscious reasoning already takes place in this first session.
However, given the above-discussed conflicting requirements, this design seems to offer the best compromise.
}

\revisionEnhancement{
\emph{Session II}: This session is intended to provide the maximum learning effect.
The subjects are instructed to ignore the sensorimotor task and are encouraged to conduct a more risky strategy, leading to more broken glasses, in order to obtain the maximum information about the underlying causal relations.
After this session, the questionnaire is applied for the second time.
}

\revisionEnhancement{
\emph{Session III}: In this session, we return to the original instructions for the sensorimotor task (as fast as possible without breaking the glass).
Since subjects are now aware of one goal of the study, the recognition of causal relations, they are assumed to follow an integrated strategy: take into account the sensorimotor instructions and, at the same time, try to identify the causal relations.
Note that these are in no sense counteracting demands.
It is due to the very design of our experiment that the best results are obtained when the causal relations are identified and used for determining the motor parameters (in particular, the optimal placement force).
It is in this session that we expect participants to reach the top level of causal sensorimotor learning.
After this session, the questionnaire is applied for the third and last time.
}

\revisionEnhancement{
Problem~\ref{item:machine_comparison}: \emph{Human-Machine Comparison.}

What would be a fair comparison of humans and machines with respect to the given experimental setup?
Ideally, a robot equipped with grasping capabilities, sensors, sensorimotor learning algorithms, and causal discovery algorithms should be brought into our experimental setup to perform the same tasks as our subjects.
This is, for various reasons, far beyond the scope of this paper.

What we can do here is to investigate how a causal representation can be learned by discovery algorithms, provided they have access to \emph{the same data as available to our human subjects.}
This is not the ideal comparison, because, for example, the force values in those data stem from our subjects, whereas our hypothetical robot would use its own strategy to adapt the motor control to the task.
However, for a first approach, this approximation appears to be reasonable.

For comparison purposes, the machine algorithms can also be fed data not directly available to our subjects, such as the breakability of glass (the force threshold), which our subjects must infer from the observable glass-ok state.
Furthermore, we can use different state-of-the-art causal discovery algorithms, such as FGES, PC, and FCI, and compare each of them to human capabilities.
Results are presented in \Cref{sect:comparison_machine}.
}

\revisionEnhancement{
\subsection{Description Levels, Causal Representations, and Causal Learning} \label{subsec:description_levels_represenation}

In this paper and in the above-described experimental paradigm, causal structures, causal representations, and causal learning methods appear on various levels, in different roles, and with different shortcuts and approximations.
We provide an overview here since a conceptually clear separation of these different aspects is essential for understanding.

\emph{Causal structures}. For the description of the ground-truth causal relations, we make use of the framework of the causal graph.
The causal structure of the most basic level of our experiment is given by the ground truth causal graph of the experimental setup, as shown in (\Cref{figure:ground_truth}).
As pointed out, the main focus of this paper is on the v-structure (common effect structure) on the upper left of the figure (also shown on the left of \Cref{figure:ground_truth_vs_representation}).
This V-structure is relevant for the human representation of the causal relations between the glass properties (weight, color, and breakability).
The complete overview of the causal structure of the experiment (including the subject) is shown in \Cref{figure:ground_truth_extended_w_sensoriomotor_representation}.
This extended causal structure results from the combination of the ground truth causal structure of the experimental setup with the (hypothetical) sensorimotor representation of the human.
It corresponds to the view the experimenter has of the situation.

\begin{figure*}[htbp]
\centerline{\includegraphics[clip, width=\textwidth]{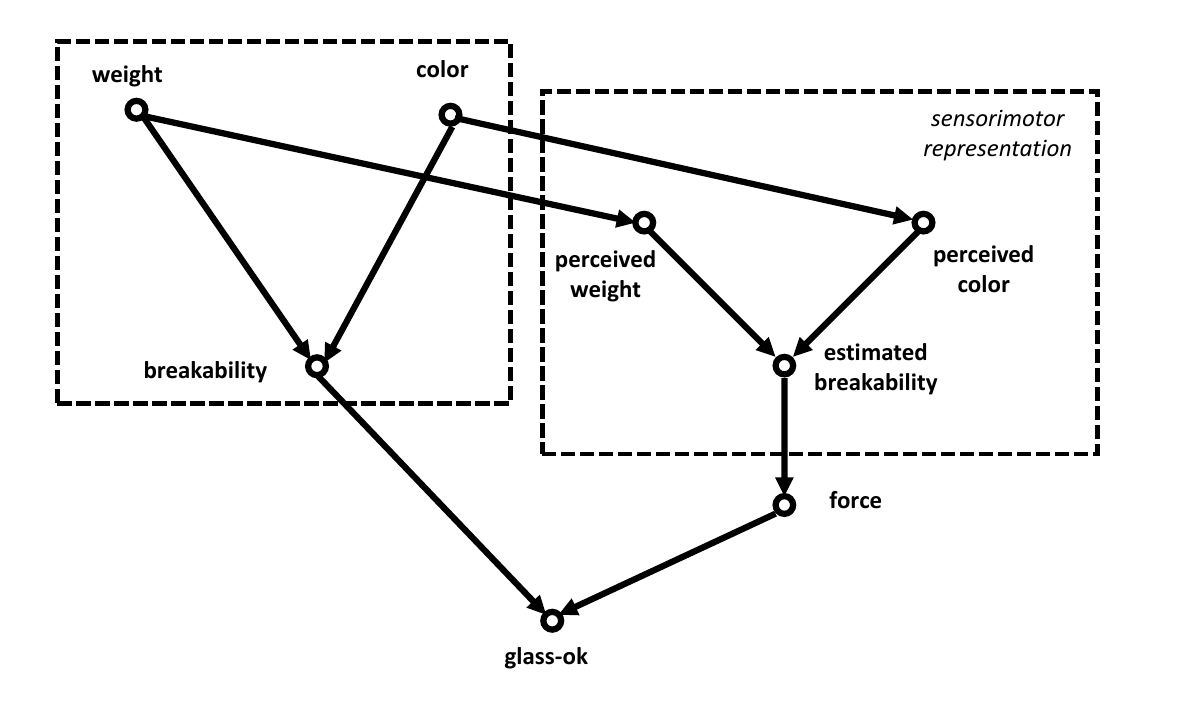}}
\caption{\revisionEnhancement{Full causal graph with ground truth causal relations of experimental setup and of the (hypothetical) sensorimotor representation.}}
\label{figure:ground_truth_extended_w_sensoriomotor_representation}
\end{figure*}

\emph{Causal representations.}  These are obtained by learning from data.
Like the ground truth causal structures, we describe them in terms of causal graphs.
We address two basic forms of causal representations in our investigation, human causal representations and "machine" representations.
Human causal representations are internal to our subjects and are assumed to exist in two variants, on a conscious cognitive level and on the sensorimotor level.
For the \emph{identification} of the respective human representations, we use two methods, questionnaires and drawings for the conscious representation, and discovery algorithms for the sensorimotor representation.
For the machine representation, no special identification is necessary; they are directly produced by the machine learning algorithms (causal discovery).
In this paper, we investigate how far our subjects can learn the ground-truth causal structure.
In the ideal limit, i.e., with perfect learning, the structure of the human causal graph will be identical to that of the ground truth.
For some of our subjects and in the different experiment sessions, the learned graph will differ from the ground truth due to missing or incorrect causal relations and directions.

\emph{Causal learning.} Corresponding to the representations are different types of learning.
For humans, there are conscious levels and sensorimotor level learning, which cannot be directly observed.
We hope to gain insight into the learning process by analyzing the development of causal representations across the three experimental sessions.
For machine learning, we use discovery algorithms, FGES, PC, and FCI.

Please note that there are two cases of learning with a special status:
a) application of machine-learning algorithms to the identification of the human causal sensorimotor representation (for reasoning see problem \ref{item:sensorimotor_causal_representation}).
b) Human-machine comparison with machine algorithms applied to human data (for reasoning see problem \ref{item:machine_comparison}).

}

\subsection{Experiment}\label{subsec:methods-dataset}

The data for the analysis were obtained from an experiment conducted with 21 healthy human adults. 
\revisionEnhancement{Since there exists neither experience with a sensorimotor task for causal structure learning nor with the use of sensorimotor data for causal discovery algorithms, the current experiment represents mainly an exploratory study.
Although we present statistical tests for some effects, the primary focus of our study is on identifying the basic properties of and relationships between causal learning at the cognitive and sensorimotor levels.
}
In this experiment, participants were required to move a glass from one plate to another as quickly as possible without breaking it (this will be further explained in the following sections and in the appendix).

The experiment is structured into three blocks, termed \textit{sessions}, each consisting of a series of 80 trials, followed by a questionnaire and a subsequent break. \revisionEnhancement{The motivation for this design is described in detail in \Cref{subsec:methods-motivation}. In the first session, termed \textit{Raw}, the participant is only told to place the glass. In this session, the participant is not aware that a questionnaire will be applied. Thereby, the participant does not pay attention to any possible relationships. As a whole, the Raw session provides a baseline for comparison. In the second session, termed \textit{Train}, the participant is allowed to experiment and explore possible strategies without any constraints or optimization targets, and is aware of the questionnaire. In the third session, termed \textit{Test}, participants were instructed to place the glass as fast as possible without breaking it. Overall, the Raw session is intended to observe the participants' naive behavior, the Train session to observe exploratory behavior, and the Test session to observe consolidated behavior.}

Each participant had an introductory, informal warm-up phase to get used to the setup. The design was verified in a pilot phase with 3 participants.

\subsubsection{Questionnaire}\label{sect:questionnaire}

After each session, the \revisionEnhancement{participant} is tasked to create a causal graph consisting of the nodes \textit{glass color, glass weight} and \textit{glass breakability} \revisionEnhancement{with qualitative (binary) relations}. \revisionEnhancement{To introduce the concept of causality, the participant is primed by an interventional definition of causality, i.e., manipulations of X propagate to Y but not the other way around \cite{pearl_2009}.}
To ease the construction, the participant may also answer binary questions regarding the existence of all possible relations. Based on the answer, \revisionEnhancement{the participant or (if necessary)} the instructor then draws the causal graph as a graphical representation of the answers. \revisionEnhancement{If the participant cannot specify the directionality of influence (e.g., the binary responses are symmetric), the edge is considered undirected. Therefore, it is not possible to give contradictory answers.}
The questionnaire is included in the appendix.

All other properties (e.g., regarding the shape or 3D model of the glass in general) are held fixed. Changing the glass weight and color is a result of careful consideration of the participants' background knowledge. The relationship between an object's weight and its breakability can be observed through everyday life. For example, fragile materials benefit from an increased width, thereby reducing overall breakability. This additional material will consequently result in a higher weight. For the color, however, there is no such simple rule. Therefore, this link between the glass' breakability and its color can be seen as \textit{more artificial} than the weight-breakability relationship. This is considered to reduce the interference of acquired knowledge throughout the experiment and prior knowledge.
After the last session, participants are asked about their optimization strategy.

\subsubsection{Hardware- and software}
To set up the VR simulation, we used the development edition of \emph{Worldviz Vizard 5.9, 64 Bit}\footnote{Available for a fee at \href{https://www.worldviz.com/vizard-virtual-reality-software}{https://www.worldviz.com/vizard-virtual-reality-software}} for \emph{python 2.7.12}. Furthermore, we used the \emph{PHANToM Premium 1.5 (High Force Model)} with the \emph{PHANToM device driver Version 5.1.7} (\cref{fig:phantom}). It is employed as a haptic device to function as both an input interface for the participants and an output device to render the force provided by the VR. \revisionEnhancement{For further specifications such as damping and stiffness see \cref{tab:appendix_phantom_specifications} in the appendix. They were chosen based on how natural they felt.} Before each session, we recalibrated the haptic device with the \emph{PHANToM configuration utility for Windows}, also \emph{version 5.1.7}. In our code for the simulation, we utilized the developer version of the \emph{sensable3e-plugin} from \emph{openHaptics}\footnote{Available for a fee at \href{https://de.3dsystems.com/haptics-devices/openhaptics}{https://de.3dsystems.com/haptics-devices/openhaptics}}.

\subsubsection{Setup}
The participant is seated in a frame with a monitor pointing downwards and mounted right above the head. The haptic device is then placed in front and in reach of the test subject. To make the content of the monitor available, an angled mirror is placed directly in front of the subject. The mirror thus also hides the participant's hand. Note that the frame blocks the participant's peripheral view due to its construction.

\subsubsection{Trial design and causal relations}
\label{subsubsec:experiment-trial-design-causal-relations}

\begin{figure}
    \centering
    \begin{subfigure}[t]{0.4\textwidth}
        \centering
        \includegraphics[width=\textwidth]{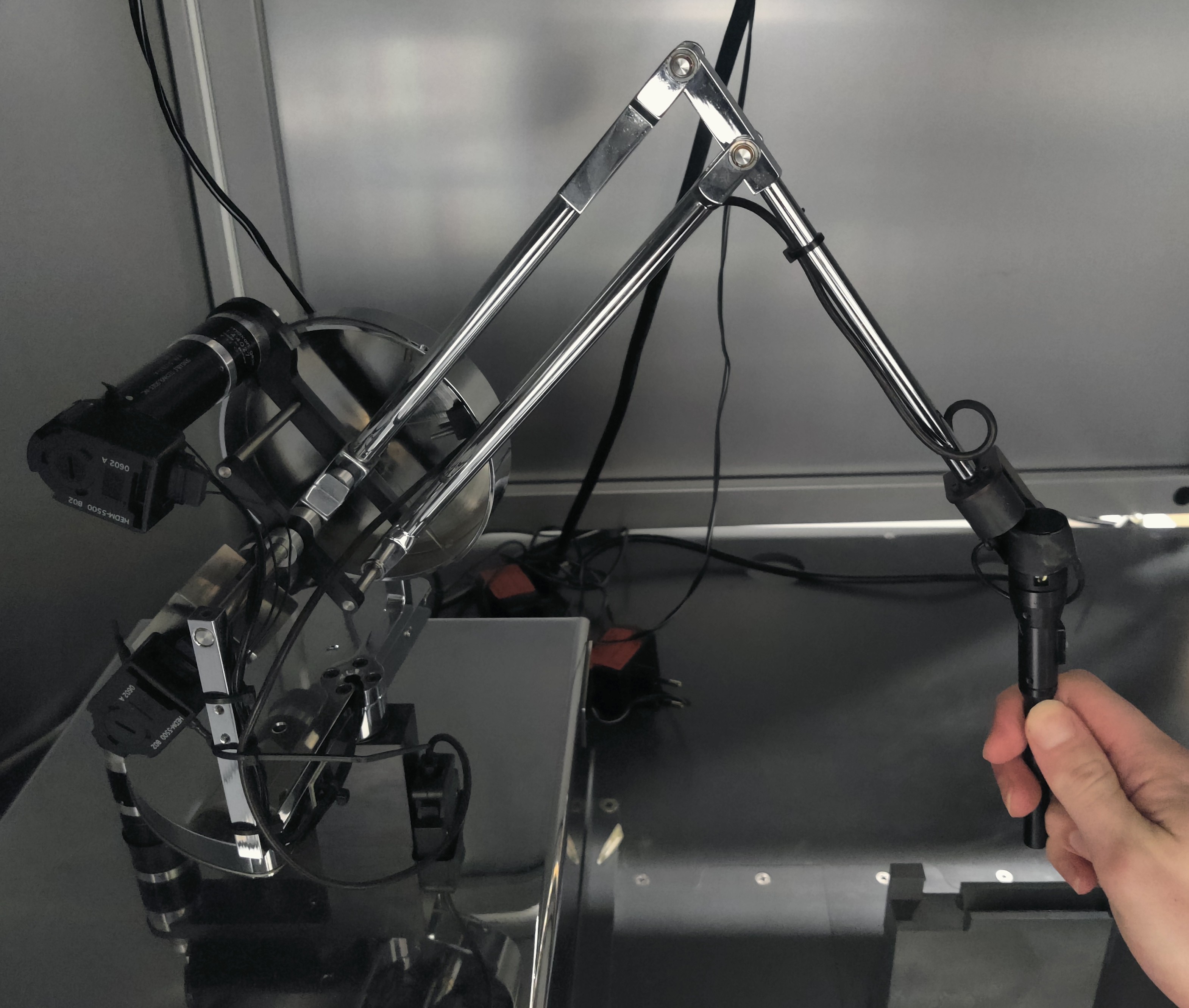}
        \caption{PHANToM haptic device used in the experiment}
        \label{fig:phantom}
    \end{subfigure}
    \hfill
    \begin{subfigure}[t]{0.5219\textwidth} 
        \centering
        \includegraphics[width=\textwidth]{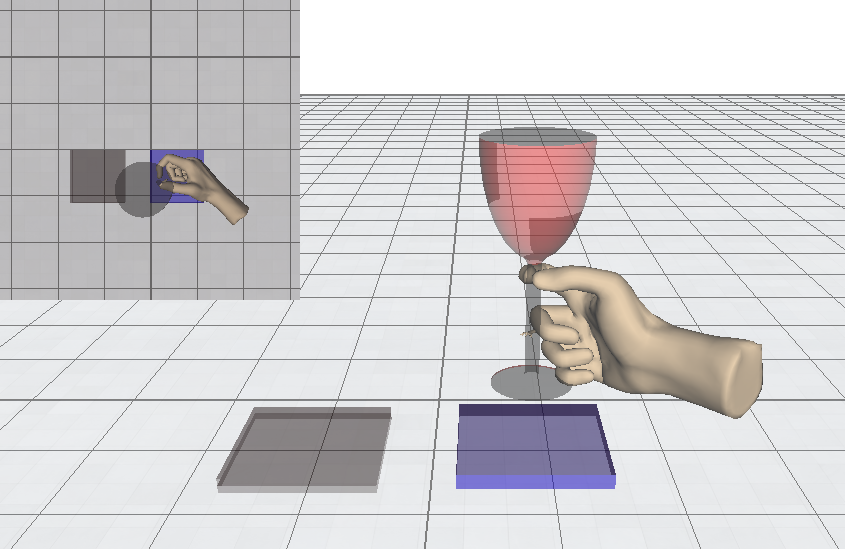}
        \caption{VR-environment of the study \revisionEnhancement{with t}wo lower plates and a bar in the upper half of the screen \revisionEnhancement{(not shown)}. The start plate appears blue, indicating that it needs to be touched next. In the upper left corner is a window that displays the top view, making it easier to locate the glass.}
        \label{fig:vr_env}
    \end{subfigure}
    \caption{Experiment setup}
\end{figure}

The VR setting consists of two plates, an indicator bar, and a closed hand holding a glass of wine (see \cref{fig:vr_env}). As described earlier, a haptic device will be used to control the glass by the user. It renders the force applied to the glass by the weight, either one of the two plates or the floor. A trial consists of placing the glass on the right plate and moving it to the left plate while touching the bar at the top at some point in between. 

The subject is instructed not to break the glass and to move it from the starting plate to the end plate as quickly as possible. The glass will break if the haptic device registers a force above a given threshold, i.e., the variable \textit{Glass-OK} in the causal model becomes zero (see \cref{figure:ground_truth}). The force threshold (i.e., the breakability of the glass) is calculated by

\begin{equation}
    \label{eq:causal_force_threshold_not_app}
    f_{\mathit{force\_threshold}}(\mathit{color}, \mathit{weight}) = (2.5\cdot \mathit{weight} + \mathit{color\_offset}(\mathit{color}))\cdot 0.8241
\end{equation}

with 

\begin{equation}
    \label{eq:causal_sampled_features_not_app}
    \begin{array}{lcl}
    \mathit{color} & \sim & \mathcal{U}(\set{\text{red, green, blue}}) \\
    \mathit{weight} & \sim & \mathcal{U}([0.2616, 1])
\end{array}
\end{equation}

and color specific offsets with the following order: $\textrm{blue} > \textrm{green} > \textrm{red}$. Therefore, on average, blue is the most stable one and red the least stable. 

The calculation of the breakability force threshold of the glass forms a collider in the causal graph. This form of interaction is deliberate, as collider/common effect structures can be identified based solely on correlational or observational data. Other forms of interaction (chain, confounder/common cause structure) leave the same trace in the data, so they are indistinguishable from one another. This is why they are in the same \emph{Markov Equivalence Class (MEC)} while colliders have a different MEC (see e.g.\cite{spirtes2001causation}). 

A more detailed explanation of the trial design and the causal model can be found in the appendix.

\subsection{Data preparation and analysis}\label{subsec:methods-data_analysis}

For each participant, there are 3 datasets, one for each session, with 80 samples each. Additionally, there is a dataset with 240 samples, composed of the concatenation of all three. This dataset resembles the experience of the participants after the experiment. \revisionEnhancement{Within each dataset, there are three continuous features and one categorical feature, making a total of four:}

\begin{enumerate}
    \item \textit{glass\_weight} - continuous.
    \item \textit{glass\_color} - categorical.
    \item \textit{force\_value} - continuous. \revisionEnhancement{The rendered contact force when the glass is placed on the plate.}
    \item \textit{force\_threshold} - continuous. The breakability of the glass.
    
\end{enumerate}

On the datasets with 240 samples, the causal discovery algorithms PC, FCI, and FGES are applied.
PC and FCI are both constraint-based algorithms that use conditional independence tests to identify the skeleton and all collider structures within the data. The FGES algorithm, on the other hand, is score-based. This class of algorithms repeatedly score, compare, and evolve candidate graphs by using a criterion such as the BIC score (see \cite{spirtes2001causation, kalainathan2022structural} for an overview). 

For PC and FCI, the conditional-gaussian likelihood ratio test (CG-LRT, \cite{andrews2018CGBIC}) and for FGES, the degenerate gaussian BIC-score (DG-BIC, \cite{andrews2019DGBIC}) are used. For the implementation of the algorithms, we used the CLI of the \emph{Tetrad toolbox for causal discovery}\footnote{\href{https://github.com/bd2kccd/causal-cmd}{https://github.com/bd2kccd/causal-cmd}} (\cite{ramsey2018tetrad}). For further analyses of the results, we used the causal-learn Python package\footnote{\href{https://github.com/py-why/causal-learn}{https://github.com/py-why/causal-learn}} (\cite{zheng2024causal}). \revisionEnhancement{Each algorithm was run 60 times with bootstrapped data, and the resulting graphs are combined such that we keep the highest frequency edge. For an overview over all hyperparameter used for Tetrad see \cref{tab:appendix_hyperparams-pc_fci,tab:appendix_hyperparams-fges} in the appendix}

\revisionEnhancement{
\subsection{Quantification of causal strength on the sensorimotor level}\label{subsec:methods-causal-strength-measure}

Assuming a causal structure $weight \rightarrow force \leftarrow color$, the strength of the effect of the glasses' $weight$ and $color$ on the contact $force$ exerted by the participants can be measured using the conditional mutual information~\citep{janzing2013mutualinformation}. The conditional mutual information \cite{Cover1991mutualInformation}, defined as:

\begin{equation}
    I(weight;force|color)= \sum_{weight,force,color}^{}P(weight,force,color) \log\frac{P(weight,force|color)}{P(weight|color)P(force|color)}
\end{equation},

quantifies the information of $weight$ on $force$ given $color$. Similarly, the expression 

\begin{equation}
	I(color;force|weight)= \sum_{color,force,weight}^{}P(color,force,weight) \log\frac{P(color,force|weight)}{P(color|weight)P(force|weight)}
\end{equation},

quantifies the information of $color$ on $force$ given $weight$. Conditional mutual information is recommended as a measure of causal strength for the structure $weight \rightarrow force \leftarrow color$ over typical measures such as the Average Causal Effect (ACE) and Analysis of Variance (ANOVA)~\citep{janzing2013mutualinformation}\footnote{ACE, defined as $P(Y = y|do(X = x_1)) - P(Y = y|do(X = x_2))$ for the interventions $do(X=x_1)$ and $do(X=x_2)$\citep{pearl2016causalStatistics}, only accounts for the linear aspect of an interaction between variables (i.e., shifts in the mean), and ANOVA is not appropriate for variables with nonlinear causal influences and when there are statistical dependencies between the variables (see \citep{janzing2013mutualinformation} and references therein).}. This quantity applies to dependencies between variables of arbitrary domains (discrete and continuous), with linear and/or nonlinear interactions \cite{janzing2013mutualinformation}. The conditional mutual information was computed in \textit{R}~\citep[Version 4.5.1]{r2022} using the \textit{infotheo} package~\citep[Version 1.2.0.1]{infotheo2022}.
}

\section{Results}
Whenever we apply causal discovery to the data (i.e. \cref{sect:sensorimotor,sect:comparison_machine}), the results obtained by the FCI algorithm are representative for the PC and FGES algorithm as well. Therefore, we only report the results based on FCI in the main document. The results of the other algorithms are presented in \cref{appendix:causal_discovery_results} as part of the appendix.
\subsection{Human Causal Discovery}
Before the results of the questionnaire are displayed, it is important to note that none of the participants \revisionEnhancement{were able to report the causal direction (i.e., the orientation of the arrows) of the relations between variables}. All participants reported that they refrained from distinguishing cause from effect because more data would be needed to do so. Therefore, only the \textit{undirected relation} between variables can be analyzed further.

\begin{figure}
    \centering
    \begin{subfigure}[b]{0.25\textwidth}
        \centering
        \includegraphics[width=\linewidth]{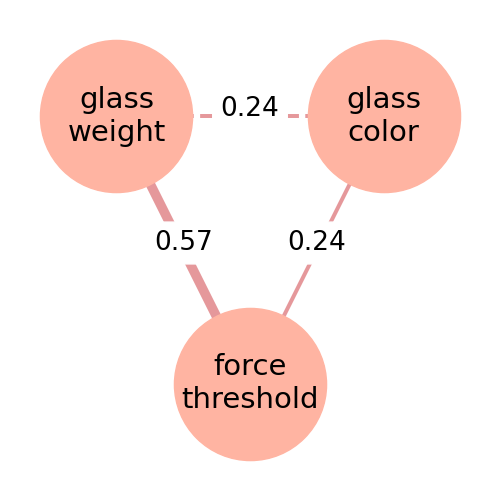}
        \caption{Session \revisionEnhancement{Raw}}
        \label{fig:results-questionnaire_answers_raw}
    \end{subfigure}
    \hfill
    \begin{subfigure}[b]{0.25\textwidth}
        \centering
        \includegraphics[width=\linewidth]{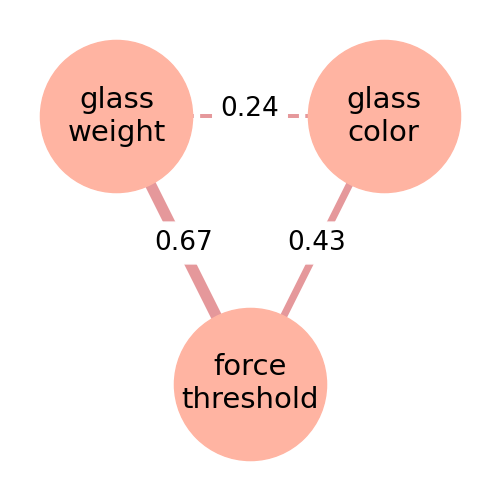}
        \caption{Session \revisionEnhancement{Train}}
        \label{fig:results-questionnaire_answers_train}
    \end{subfigure}
    \hfill
    \begin{subfigure}[b]{0.25\textwidth}
        \centering
        \includegraphics[width=\linewidth]{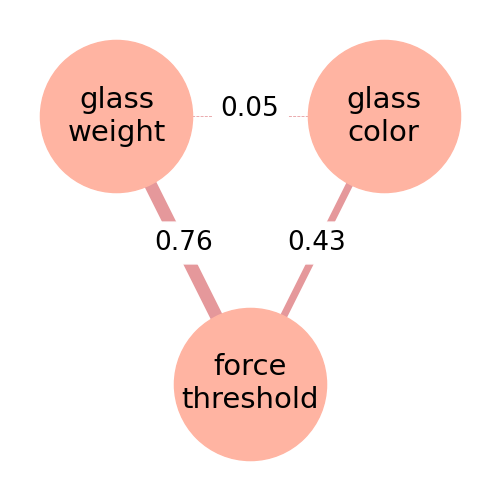}
        \caption{Session \revisionEnhancement{Test}}
        \label{fig:results-questionnaire_answers_test}
    \end{subfigure}
    \centering
    \caption{Aggregated results of the questionnaire. The thickness of the lines corresponds to \revisionEnhancement{the fraction of} how many participants identified the relation between the connected variables. \revisionEnhancement{This fraction is also shown on the given lines.} The dashed connection indicates that this connection is erroneous.}
    \label{fig:results-questionnaire_answers}
\end{figure}

\Cref{fig:results-questionnaire_answers} shows the aggregated results of the questionnaire. The thickness of the connection corresponds to the number of participants who identified a relation between the connected variables. The dashed line indicates an erroneously identified connection.

\begin{figure*}[htbp]
\centerline{\includegraphics[clip, width=\textwidth]{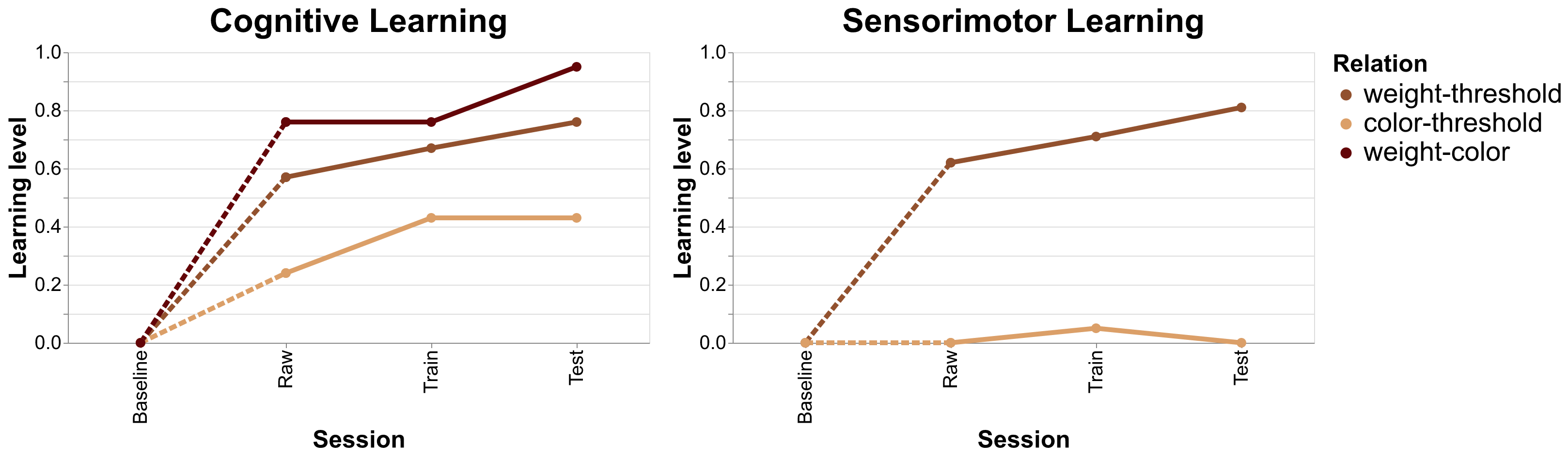}}
\caption{\revisionEnhancement{Learning on the conscious cognitive level (left) and sensorimotor level (right) from the initial state of "no information" across the three experiment sessions (percentage correct responses for each edge). The dashed lines indicate the assumed change from baseline to the first observed response.}}
\label{figure:results-cogni_sensori_learning_curves}
\end{figure*}

\revisionEnhancement{\Cref{figure:results-cogni_sensori_learning_curves} shows the same information with respect to the learning effect from the initial state before the experiment (here assumed with "no causal information" = 0\%) across the three experiment sessions.}

Considering the assessments of the humans in their entirety, the structure identification becomes more accurate over the course of the experiment. The connection between the weight of the glass and its breakability was the most prominently identified one. In the first session, $57\%$ of participants suspected it, $67\%$ in the second session, and $76\%$ in the last session. The relationship between the glass color and its breakability seems to be harder to identify, as only $24\%$ recognized the relation in the first session, which increased to $43\%$ in the subsequent sessions. 

After the first session, some participants noted that they expected the weight of the glass to be connected to its color. This suggests a strong bias towards establishing a causal relationship between the two properties.
Accordingly, we observed that $24\%$ of the participants drew a connection between the two variables in the test after the first two sessions. However, the percentage drops to less than $5\%$ after the last session, when more empirical data for their decision has been gathered by the subjects.

\subsection{Sensorimotor}\label{sect:sensorimotor}

\begin{figure}
    \centering
    \begin{subfigure}[b]{0.25\textwidth}
        \centering
        \includegraphics[width=\linewidth]{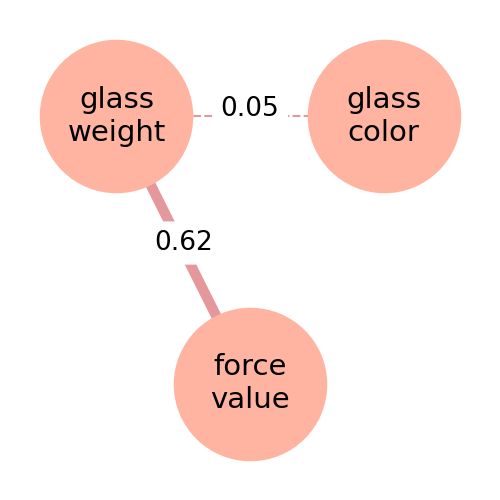}
        \label{fig:results-sensorimotor_causal_representation_raw}
    \end{subfigure}
    \hfill
    \begin{subfigure}[b]{0.25\textwidth}
        \centering
        \includegraphics[width=\linewidth]{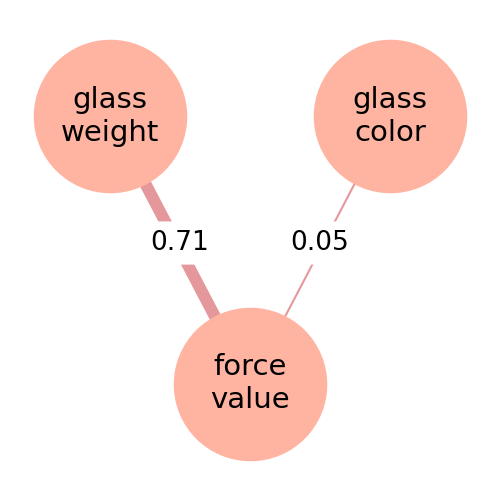}
        \label{fig:results-sensorimotor_causal_representation_train}
    \end{subfigure}
    \hfill
    \begin{subfigure}[b]{0.25\textwidth}
        \centering
        \includegraphics[width=\linewidth]{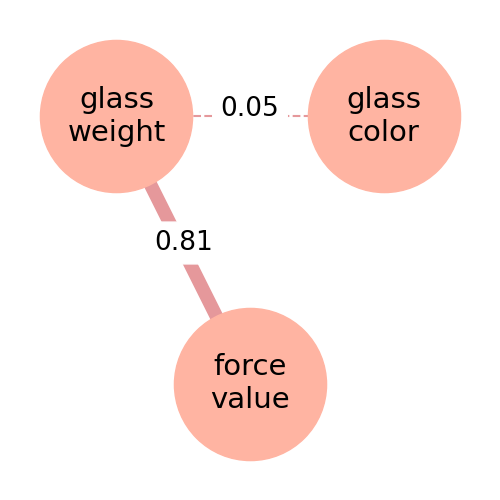}
        \label{fig:results-sensorimotor_causal_representation_test}
    \end{subfigure}
    \centering
    \caption{\revisionEnhancement{Sensorimotor causal graphs identified by the FCI causal discovery algorithm (sessions from left to right: raw, train, test)}}
    \label{figure:results-sensorimotor_causal_representation}
\end{figure}

During interrogation, the participants reported sensorimotor optimization strategies that were completely agnostic to the properties of the glass (e.g., optimizing muscle control and movement path, reducing the distance to move, etc.).
\revisionEnhancement{
Indeed, the moved distance and speed in general increased over the duration of the experiment.
The self-reported stimulus-agnostic optimization is surprising because the average gap between the applied force and the breakability threshold reduces from the first to the last session. Additionally, the graph of the conscious cognitive causal representation (\Cref{fig:results-questionnaire_answers}) shows that the majority of subjects (76\%) clearly recognized a causal dependency between the breakability of the glass and its weight, and 43\% recognized this dependency for color as well.
We thus examine the sensorimotor level in more detail to determine whether it is really agnostic to the causal dependencies.

We derived the causal graph of the sensorimotor representation by application of discovery algorithms to the recorded experimental data.
(The reasoning and the method for this is described in \Cref{subsec:methods-motivation},~Problem~\ref{item:observable_sensorimotor_representation}.)
The resulting causal graph (FCI algorithm) is shown in \Cref{figure:results-sensorimotor_causal_representation} and the corresponding learning curves are shown in \Cref{figure:results-cogni_sensori_learning_curves}. See \cref{appendix:causal_discovery_sensorimotor} in the appendix for results of the PC and FGES algorithms.
}

\revisionEnhancement{
The results provide clear evidence that the sensorimotor level is \emph{not} completely agnostic to the causal relations.
There is a causal link (0.81) between the glass weight and the force value rendered during the contact of the glass and the plate.
This is even slightly higher than the corresponding value on the conscious cognitive level (0.76).
Note that since the sensorimotor causal representation is derived from data that include the rendered force value that the subjects perceive during the contact of the glass and the plate, this result also shows that subjects incorporate the causal dependency between weight and breakability\_threshold into their sensorimotor strategy, and this without conscious awareness of doing so.

The situation is different for the color-force link.
Such a link could be found on the 0.05 level in a few discovery runs, but this could also be a spurious result.
Since discovery algorithms operate on the safe side (rather than missing a weak causal dependency and reporting it erroneously as existing), we sought a more sensitive measure.
For this, we directly examined the correlations between weight and force, as well as between color and force.
}

\begin{figure}
    \centering
    \includegraphics[width=\textwidth]{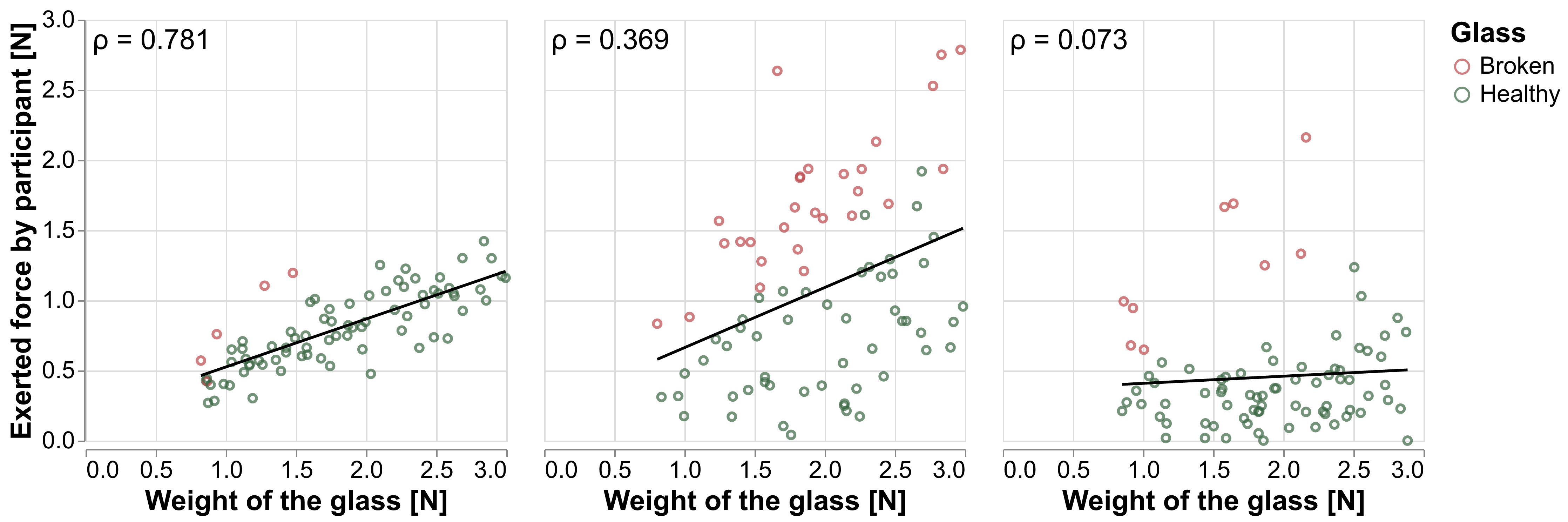}
    \caption{Examples for different levels of \revisionEnhancement{linear} correlation \revisionEnhancement{(shown in upper left corner)} of the glass weight and the exerted force by the participant on the glass (both in N). The data has been recorded in the last session of the experiment and stem from three different participants. Each circle represents a unique trial, and the black line indicates a linear regression of the force on the weight.}
    \label{fig:results-weight_force_relationship_examples}
\end{figure}

\begin{figure}
    \centering
    \includegraphics[width=\linewidth]{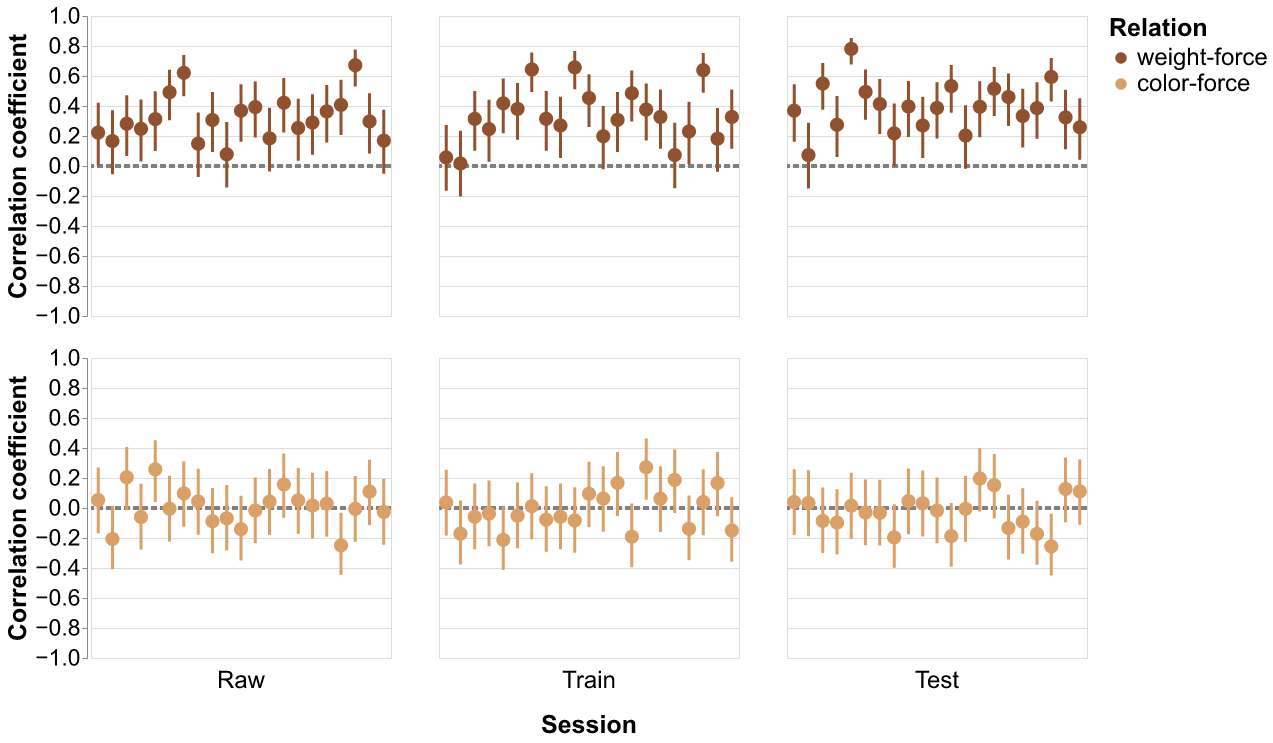}
    \caption{\revisionEnhancement{Linear correlations for the causal weight-force (top) and color-force (bottom) relations. The confidence intervals are approximated using Fisher's z-transformation. The gray lines indicate zero correlation.}}
    \label{figure:force_correlations}
\end{figure}

This closer analysis of the causal weight-force link is illustrated in \cref{fig:results-weight_force_relationship_examples} and \cref{figure:force_correlations} (top).
\revisionEnhancement{
If the sensorimotor system were to completely ignore the influence of the glasses' weight, the correlations should be zero.
In contrast, almost all correlations are found to be greater than zero, including the bounds of the confidence intervals calculated using Fisher's z-transformation.
}
\revisionEnhancement{Furthermore, a two-sided one-sample t-test on the change in linear correlation from the first to the last session indicated a significant difference from zero with $\alpha = 0.05$ ($t(20)=2.253, p=0.035$). The mean change was positive ($\overline{\mu} = 0.073$), indicating an increase in causal coupling.}
\revisionEnhancement{
This implies that further learning of the weight-force relationship occurs on the sensorimotor level during the experiment.
}
Please also note that no negative correlation \revisionEnhancement{has been observed} (which would correspond to an increased caution when placing a heavy glass). Additionally, the difference between the rendered force and the force threshold needed to break the glass is reduced in the last session compared to the first. 

\revisionEnhancement{
In a further step, we applied correlation analysis to the causal relation between color and force, which was unclear in the sensorimotor causal graph obtained by discovery analysis (\Cref{figure:results-sensorimotor_causal_representation}).
Linear correlations are shown in \Cref{figure:force_correlations}.
They were dominated by noise, showing negative and positive correlations, which signals that the sensorimotor system had difficulties in identifying the correct linear ordering of the color-breakability mapping as defined in \Cref{subsubsec:experiment-trial-design-causal-relations}.
}

\revisionEnhancement{
However, this does not exclude that some information is taken into account, e.g., on the color which most or least enhances sturdiness, while the other two colors are unsystematically mapped to force.
We therefore sought a measure that is not based on linear relations or correct ordering.
Such a measure is given by the mutual information.
In this context, it is interesting to note that mutual information has been suggested as the preferable measure for quantifying the causal influence of one variable on another~\citep{janzing2013mutualinformation}.
}

\begin{figure*}[htbp]
\centerline{\includegraphics[clip, width=0.7\textwidth]{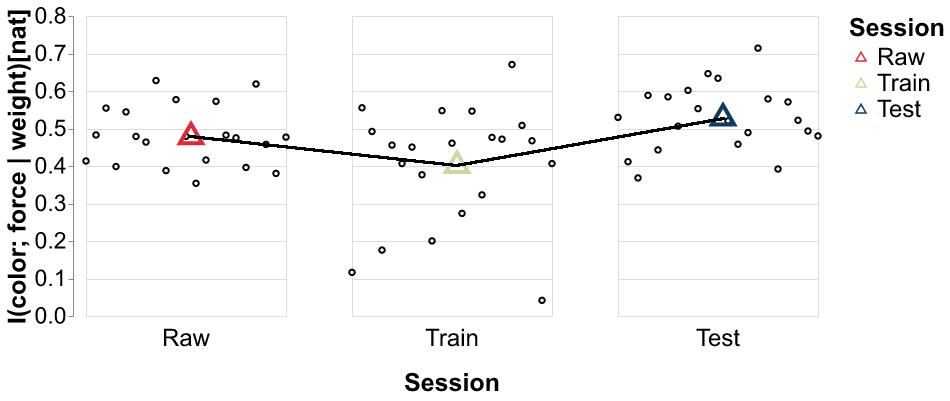}}
\caption{\revisionEnhancement{Mutual information for the causal color-force relation.}}
\label{figure:color-force_mutual_information}
\end{figure*}

\revisionEnhancement{
The mutual information values for the color-force relations across subjects and sessions are shown in \Cref{figure:color-force_mutual_information}. For the interpretation, recall that $I (color; force| weight ) = 0$ indicates that $color$ and $force$ are conditionally independent given $weight$ (i.e., $color$ does not influence $force$ when $weight$ is known)~\citep[p. 35]{Cover1991mutualInformation}. 
\Cref{figure:color-force_mutual_information} shows three interesting facts:
First, the observed $I (color; force| weight )>0$ indicates that there is a causal influence of color on force, i.e., the sensorimotor system takes color into account when determining the motor parameters. Second, although the sensorimotor system does not find the right solution within the duration of the experiment, it learns and improves, as evidenced by the increase from the first (Raw) to the last session (Test) of the experiment.
Finally, some low mutual information values appear in the second session (Train), which indicate a low influence of $color$ on $force$.
Note that this is the session for which subjects have been encouraged to test out different exploratory behaviors, irrespective of how many glasses will break.
For example, if a subject decides to test what happens if they will use the same force in the next 10 trials, irrespective of the glass properties and of how many glasses break, then the force will be become "decoupled" from the setup and the mutual information will go down, approaching the value $I (color; force| weight ) = 0$ (i.e., approaching conditional independence).
}

\subsection{\revisionEnhancement{Comparison of the causal relations on conscious cognitive level and on the sensorimotor level}}
\label{sect:comparison_cognitive_sensorimotor}
\begin{figure*}[htbp]
\centerline{\includegraphics[clip, width=0.5\textwidth]{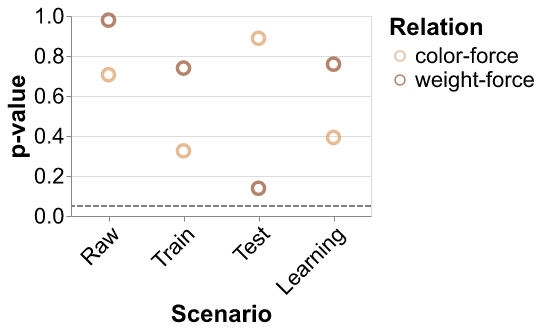}}
\caption{\revisionEnhancement{P-Values of Alexander-Govern tests performed with null-hypothesis of equal means. The conditional mutual information values of participants are analyzed, who are separated into groups depending on whether they perceived the given causal relation (indicated by color). In "Learning", the change of the CMI value from session Raw to session Test is considered.}}
\label{fig:results-statistical_tests_perceived_sensori}
\end{figure*}

\revisionEnhancement{To compare the cognitive and sensorimotor level, we split the participants into two groups per session, depending on whether they identified the relation of interest. In addition to the split per session, we calculate the change of conditional mutual information from the first to the last session per participant. We compared them using the conditional mutual information values.\\
The Alexander-Govern tests indicate no significant difference between participants who did perceive and did not perceive the causal relation in any scenario tested (\cref{fig:results-statistical_tests_perceived_sensori} shows the p-values).}

\begin{figure}
    \centering
    \includegraphics[width=0.9\linewidth]{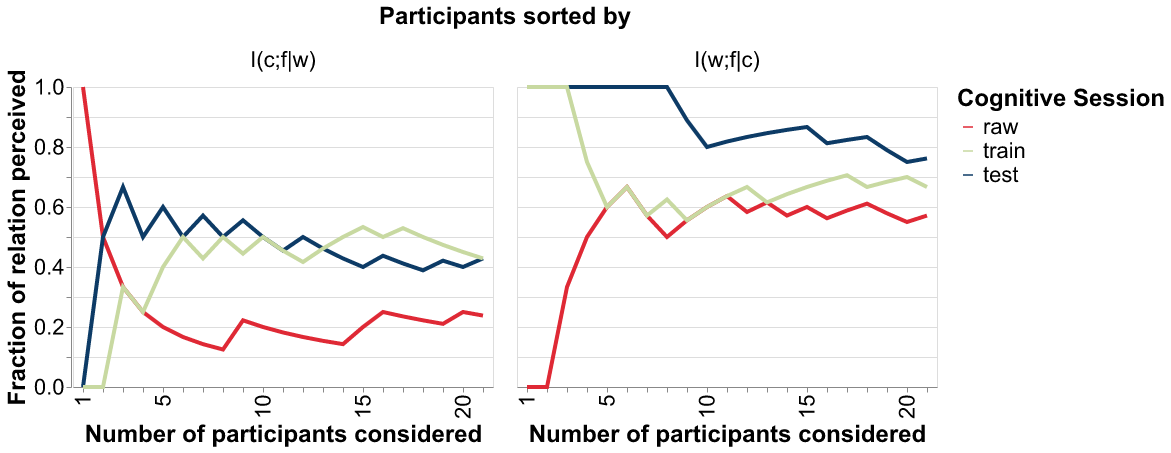}
    \caption{\revisionEnhancement{Fraction of participants that perceived each causal relation (color-breakability left, weight-breakability right) as a function of top-ranked participants considered for each session. The ranking is based on the mutual information of the applied force by the participant and the given stimulus, conditioned on the other stimulus. The x-axis shows the number of participants included.}}
    \label{fig:results-cognitive_sensorimotor_rank_plot}
\end{figure}

\revisionEnhancement{While the statistical tests did not yield statistically significant results, subtle interactions can be found in the data worth exploring further. To complement the statistical tests, we employ alternative, more exploratory measures.\\
\Cref{fig:results-cognitive_sensorimotor_rank_plot} depicts the interplay of the cognitive- and sensorimotor-level. The plot shows the fraction of participants who judged a causal relationship between glass color and breakability (left) and glass weight and breakability (right), as a function of the number of top-ranked participants included. The ranking is based on the mutual information between stimulus features (color, weight) and the contact force of the glass on the surface, given the other stimulus feature. The x-axis shows the number of participants considered, starting with those exhibiting the strongest sensorimotor-stimulus association, and the y-axis indicates the proportion of these participants who identified the respective causal relationship based on the questionnaire. The color indicates the session.\\
The left plot indicates that the fraction of participants who perceive the causal relation between color and breakability, as well as the mutual information between color and force, is independent of each other, regardless of the session. However, in the right plot, the mutual information (specifically in the test session) is highest in participants who also identified the causal relation, thus indicating a communication between the sensorimotor and the cognitive level.}

\begin{figure}
    \centering
    \includegraphics[width=0.8\linewidth]{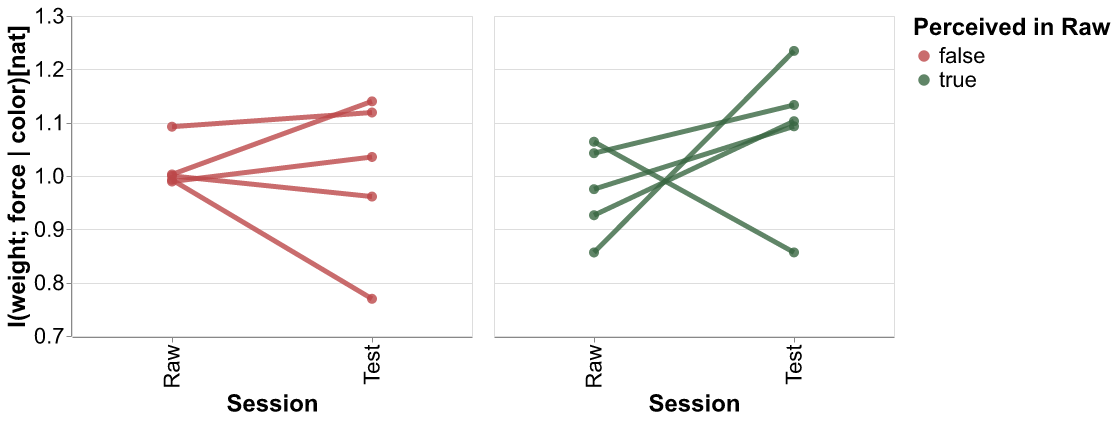}
    \caption{\revisionEnhancement{Change of sensorimotor mutual information from first to last session conditioned on the conscious cognitive recognition of the weight-breakability relation in the first session.}}
    \label{fig:results-sensorimotor_conscious_datalines}
\end{figure}

\revisionEnhancement{Next, and focusing on the weight-force relation, we considered the participants with low conditional mutual information values and compared the gain from the first to the last session. \Cref{fig:results-sensorimotor_conscious_datalines} shows that the gain is higher in participants who perceived the causal relation compared to non-perceiving ones. In the latter group, the mutual information is mostly stable.}

\subsection{\revisionEnhancement{Comparison With Machine Algorithms}}
\label{sect:comparison_machine}

\revisionEnhancement{
As described in \Cref{subsec:methods-motivation},~Problem~\ref{item:machine_comparison}, a real machine comparison using a robot in the experiment is not possible.
Therefore, we apply the machine algorithms for causal discovery to the data from the experiment.
We provide the same data as used for and being generated by each of the subjects, organized into 3 sessions with 80 trials each.

We begin with an idealized test, in which we grant the algorithms access to the ground-truth variables (weight, color, and breakability threshold).
Note that this is a somewhat unfair comparison for humans, as the latter variable is not directly observable to them and they have no 100\% perfect perception of weight and color.
The main motivation for this test is (a) to provide a baseline on the convergence and robustness properties of the machine algorithms under ideal conditions and (b) to get an idea of what could potentially be learned under the conditions of our experiment.
}

\begin{figure}
    \centering
    \begin{subfigure}[b]{0.3\textwidth}
        \centering
        \includegraphics[width=\linewidth]{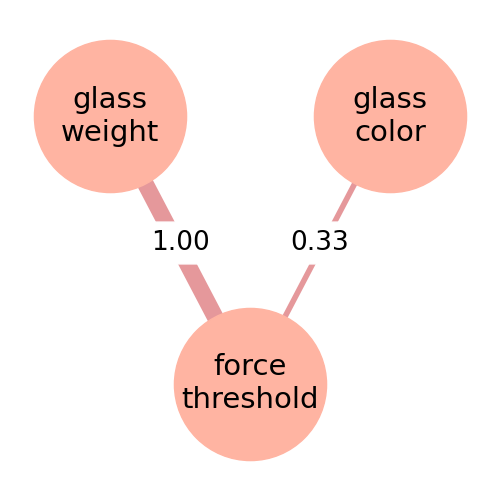}
        \label{fig:results-machine_discovery_idealized_raw}
    \end{subfigure}
    \hfill
    \begin{subfigure}[b]{0.3\textwidth}
        \centering
        \includegraphics[width=\linewidth]{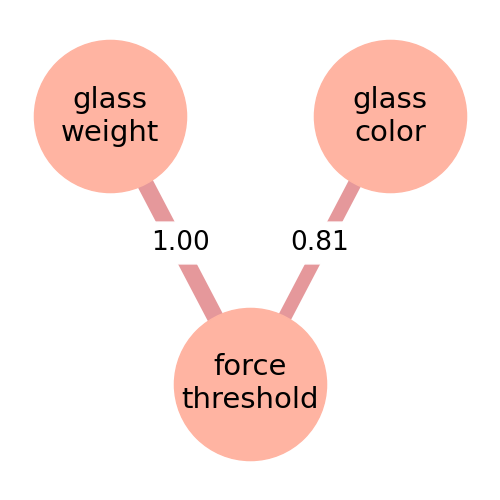}
        \label{fig:results-machine_discovery_idealized_train}
    \end{subfigure}
    \hfill
    \begin{subfigure}[b]{0.3\textwidth}
        \centering
        \includegraphics[width=\linewidth]{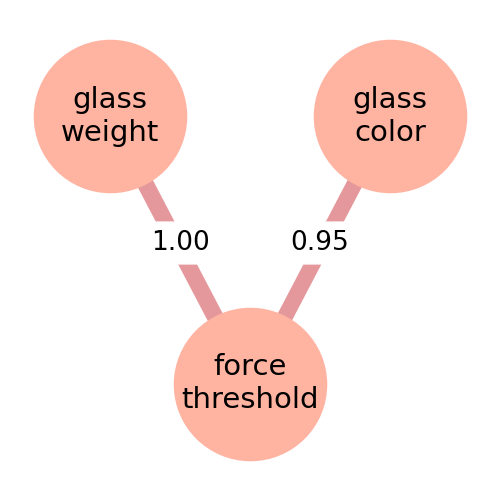}
        \label{fig:results-machine_discovery_idealized_test}
    \end{subfigure}

    \centering
    \begin{subfigure}[b]{\textwidth}
        \centering
        \includegraphics[width=\linewidth]{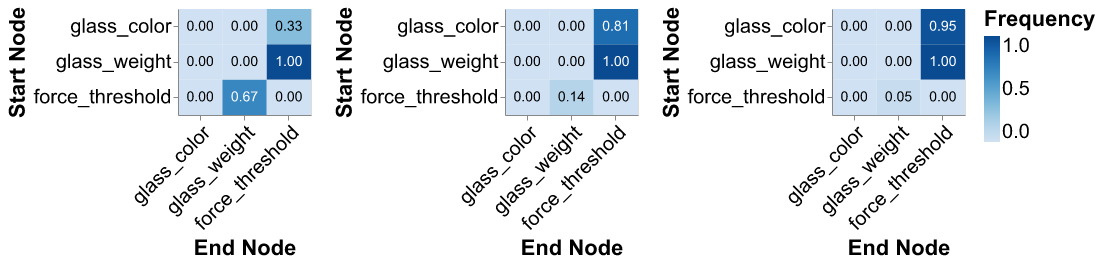}
        \label{fig:results-machine_discovery_idealized_directions}
    \end{subfigure}
    \caption{\revisionEnhancement{Causal machine learning with discovery algorithms across the three sessions with an idealized variable set. Upper part: Causal connectivity. Lower part: Causal directions.}}
    \label{figure:results-machine_discovery_idealized}
\end{figure}

\revisionEnhancement{
The results of this test for FCI are shown in \Cref{figure:results-machine_discovery_idealized}.
The causal graph is shown after learning from the data from the first session, from the combined data from the first and second sessions, and after learning from all three sessions.
The results are obtained from data of single-subject runs and then averaged over subjects. See the appendix \cref{appendix:causal_discovery_pc_idealized,appendix:causal_discovery_fges_idealized} for the results for PC and FGES algorithms.

After the first session, the causal connection weight~--~breakability\_threshold is learned perfectly, but the causal direction is unstable, as that can only be established in conjunction with a third variable to form a V-Structure.
The connection color~--~breakability\_threshold is recognized in 33\% of the runs and forms this V-Structure to establish the causal direction. The rate of identification is higher than the 24\% of the human subjects.
The connection weight~--~color is correctly recognized as non-existing, in contrast to it being erroneously detected by 24\% of the humans.

During the next two sessions, the learned causal graph stabilizes. 
After incorporating all three sessions, the learned causal graph accurately represents the ground truth almost perfectly. In comparison to the development and quality of the machine representation in quantitative detail to those learned by the participants (\Cref{fig:results-questionnaire_answers} and \Cref{figure:results-sensorimotor_causal_representation}), the machine seems superior. However, this comparison is idealized for machine learning and unfair insofar as humans cannot directly observe the breakability threshold.
}

\revisionEnhancement{

Therefore, we investigate in the next step what machine algorithms can learn if they can access only the variables observable to the human subjects.
Instead of the non-observable breakability\_threshold, we now make use of the two observable variables force and glass-OK.\\
It is essential to understand the subtle but important difference in the application of discovery algorithms in \Cref{sect:sensorimotor}. For the analysis of the human causal representation, we wanted to know what the subjects have \emph{actually} learned; for the human-machine comparison, we want to know what could \emph{potentially} be learned from the data. The analysis of the human sensorimotor causal representation has thus been limited to only those three variables whose causal relations are relevant for the human causal representation (see \Cref{figure:ground_truth_vs_representation}). In particular, the target variable regarding the causal representation on the sensorimotor level is the force (as a proxy for the estimated breakability). The outcome, i.e., the Glass-OK status of the glass \emph{after} application of the result of the sensorimotor computation, is not part of the causal sensorimotor representation. It could of course be used by the sensorimotor system in its learning but we are not interested in the question of \emph{how} we causally describe the learning mechanism of the sensorimotor system but we want to describe the causal sensorimotor representation which \emph{results} from the learning.\\
This differs for the question being considered now: what can potentially be learned from the data (in particular, by a machine algorithm)? For this, it is essential to include the Glass-OK status in the analysis.

The results of applying causal discovery (FCI) to the four variables are shown in \Cref{fig:machine-representation-observable-4var}. Similar to the idealized set of variables, the algorithms are applied to the cumulative combined data from the three sessions of one participant, and the results are averaged over all subjects. See the appendix, \cref{appendix:causal_discovery_comparison_realistic} for results of the PC and FGES algorithms.

\begin{figure}[h]
    \centering
    \begin{subfigure}[b]{0.3\textwidth}
        \centering
        \includegraphics[width=\linewidth]{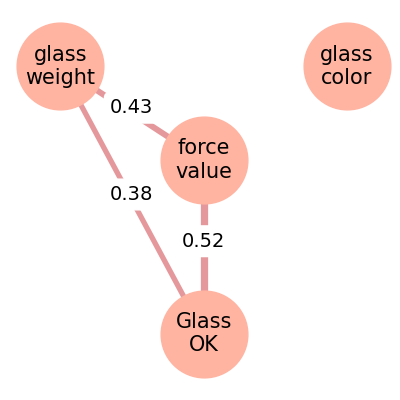}
    \end{subfigure}
    \hfill
    \begin{subfigure}[b]{0.3\textwidth}
        \centering
        \includegraphics[width=\linewidth]{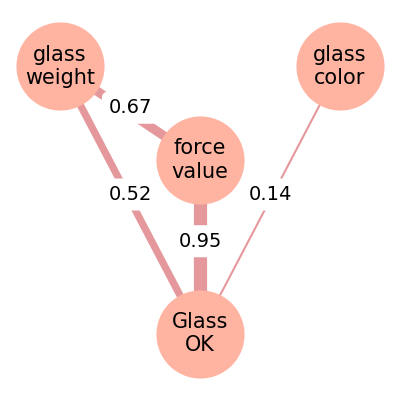}
    \end{subfigure}
    \hfill
    \begin{subfigure}[b]{0.3\textwidth}
        \centering
        \includegraphics[width=\linewidth]{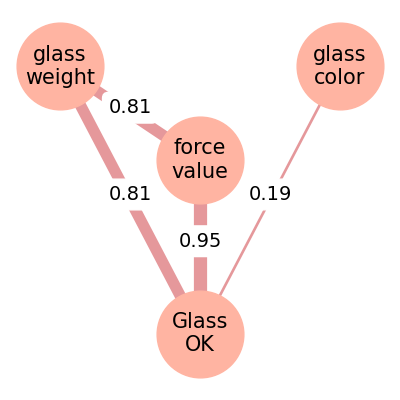}
    \end{subfigure}
    \centering
    \caption{\revisionEnhancement{Causal relations discovered by the FCI algorithm from the observable data available to humans in the experiment (datasets contain sessions raw (left), raw, train (center),  raw, train and test (right)}}
    \label{fig:machine-representation-observable-4var}
\end{figure}

The glass\_weight~-~glass\_ok relation has been recovered in 81\% of runs based on the combined data of all sessions (right), whereas the glass\_color~-~glass\_ok relation is identified in 19\% of runs. There was no error regarding the link between glass\_color and glass\_weight.
Therefore, the performance of the algorithms is comparable to that of our participants for the influence of weight on the breakability (76\%) as well as the independence of glass\_weight and glass\_color (0.05\%) and worse for the influence of the color on the breakability (43\%, see \cref{fig:results-questionnaire_answers}).
Additionally, the direct link between force\_value and the outcome of the experiment is recovered in almost all runs (95\%). See \cref{sect:comparison_with_machine_algorithms} for details about the interpretation of (absent) relations, including force\_value.
}

\section{Discussion}\label{sect:discussion}
\subsection{\revisionEnhancement{Experimental Paradigm}}
\label{sect:experimental_paradigm}
\revisionEnhancement{
In our view, there are several novel aspects of this study:
First, unlike many studies on causal structure learning that use abstract settings, we utilize a naturalistic setup based on sensorimotor object place tasks. 
Second, to the best of our knowledge, this is the first paradigm developed to measure causal structure learning at the sensorimotor level. 
Third, our setup enables a comparison between causal learning on conscious cognitive and sensorimotor levels, facilitated by our experimental design that ensures both levels identify the same ground-truth causal structure.
}
\subsection{\revisionEnhancement{Conscious Cognitive Level}}
\label{sect:conscious_cognitive_level}

\revisionEnhancement{
The results indicate that humans can identify the causal V-structure in our experimental setup through observational learning, achieving 76\% weight-breakability, 43\% color-breakability, and 95\% weight-color independence. Previous studies in more abstract contexts suggested that observing covariation often fails to lead to accurate causal structure identification~\citep{lagnado2002learning,steyvers2003inferring, lagnado2004advantage,lagnado2007beyond}. Our findings present a more hopeful view of human competence in structure identification. It is plausible that the handling of causality and probability varies based on how problems are presented~\citep{meder2014statisticalThinking}. While artificial setups are preferred to avoid prior experience biases, future research should explore more natural settings.

}

\revisionEnhancement{What is unexpected is that our subjects had substantial difficulties in recognizing the direction of causality in our experiment. We expected that by conventional reasoning one direction would be preferred (with inter-individual differences)\footnote{E.g. More weight indicates more material and therefore less breakability}}
\revisionEnhancement{The inability or reluctance} to distinguish cause from effect has been reported on other occasions as well.
A plausible explanation of such problems is associative reasoning, which is directionless by definition \citep{rottman2014reasoning, rehder2001causal}.
Others have reported that humans may perform local reasoning schemes by only considering two variables at a time (\cite{rottman2014reasoning, kruschke2006locally, fernbach2011asymmetries, waldmann2008causal}). 
In fact, the participants' verbalizations during the questionnaire suggest the utilization of this strategy. Furthermore, it is proposed that people sometimes may add imaginary nodes to such graphs \citep{rehder2005feature,rottman2014reasoning}. \revisionEnhancement{Therefore, an imagined hidden common cause could make potential causal relations seem like spurious correlations without direction. Removing a link for such a correlation would rely on the participant's understanding of the question.}

Another interesting observation is that some participants noted a strong prior belief that weight and color will possibly be connected in our experimental setup \revisionEnhancement{(see also role of prior)}. 
With this in mind, it is not surprising that this connection also appears in the graphs identified by the subjects after the first session, as these biases have a large impact on the (perhaps erroneous) detection of relationships. 
A recognition of correlation in data is simpler than rejecting a hypothesized relationship (\cite{waldmann1996knowledge, waldmann2001estimating,mercier2022confirmation,klayman1995varieties}, see \cite{lagnado2007beyond}).

\revisionEnhancement{We do not believe this reflects actual prior experience, as no transparent glasses were used, and the red, green, and blue glasses do not have an expected systematic relation to weight. Many types of colored glasses have varying weights. Subjects recognized color and weight, and maybe they initially expected a relation due to the experimental context. However, most were able to learn that no such relation exists. This unlearning is notable, as representations of causal links can be resistant to contradictory information~\citealp{barberia2019persistenceCausalIllusions,lagnado2006time,yarritu2015illusionCausality}.}
\subsection{\revisionEnhancement{Sensorimotor Level}}
\label{sect:sensorimotor_level}

 \revisionEnhancement{

We conducted a novel investigation of causal structure learning at the sensorimotor level  (see comments on problems 4 and 6 in sect. 3.1) through a specific task and meta-level analysis using machine causal discovery methods. The causal discovery analysis indicated that the sensorimotor system "recognized" weight as influencing breakability, while no causal link was found between color and breakability. Correlation analysis supported the weight-breakability relationship with positive correlations, though not as systematic as those at the cognitive level, and showed no evidence of a color-breakability relationship, with correlation values around zero.

However, does the sensorimotor system really completely ignore the color-breakability relation? Applying a mutual information measure showed that this is not the case. In case of complete ignorance of the relation between color and breakability, the force variable should be simply independent of color. However, non-zero mutual information was found in the first session. Additionally, since the information measures increased, the sensorimotor system was likely able to learn and optimize using the color-breakability relation during the experiment.

The system struggled to learn the correct quantitative relationship between color and breakability, as shown by its random correlation results. This difficulty arises from the complex mapping and the challenge of ordering color levels. Moreover, color has less influence on breakability than weight, based on our cautious interpretation of pre-test results. Future experiments emphasizing color could improve the understanding of this relationship. Simplifying color information from three to two levels might also help clarify the linear relationship with breakability.

Finally, it is worth noting that in sensorimotor research, a decrease in errors indicates learning. Using mutual information could offer insights into early sensorimotor learning phases and enhance causal discovery results.
}

\subsection{\revisionEnhancement{Relations Between Levels}}
\label{sect:relations_bewteen_levels}
\revisionEnhancement{

The current state of scientific opinion on this issue remains uncertain (see problem \ref{item:levels_causal_representation} in \cref{subsec:methods-motivation}). Our data indicate that information transfer between cognitive and sensorimotor levels is largely decoupled. While exploratory measures suggest a weak relationship, we have found no evidence for the direction of information flow.

If both levels share the same causal representation, subjects recognizing the causal relation should show stronger coupling in the sensorimotor domain. This could occur on the same timescale (indicating that both levels operate in parallel) or with a slight delay (indicating that one level informs the other). However, our findings suggest unrelated causal representations in early learning stages, as there is minimal overlap between those recognizing the causal relation and those showing increased stimulus-motor output association.

In comparison, the cognitive level appears to identify relations more quickly and precisely than the sensorimotor level, which faces greater demands. While the cognitive level may only need to understand qualitative relations, the sensorimotor level must determine quantitative relationships for successful task completion, making it more challenging.
}

\revisionEnhancement{

For further analysis, we need to differentiate the levels, their interrelations, and the information they represent. We focus on two basic levels: cognitive, measured by the questionnaire, and sensorimotor, assessed through sensorimotor data. While we've examined the causal representational properties of these levels, there's also an "intermediate" level related to self-reported behaviors and optimization strategies. Participants reported how they optimize their movements, which reflects their conscious perception of sensorimotor properties. This intermediate level can be viewed as a sub-level of the cognitive level.
}

This distinction of three levels now allows for the observation of an interesting pattern of dissociation. 
\revisionEnhancement{On the basic cognitive level}, the relationship between the weight and the \revisionEnhancement{breakability} of the glass can be clearly perceived and verbalized. 
This relationship also appears to be accessible at the \revisionEnhancement{basic sensorimotor level}, where participants seem to improve over the course of the experiment in placing heavier glasses with less precision without breaking them (an important possibility is that this allows for exploiting the speed-accuracy-tradeoff~\citep{burdet1998quantization}).
However, when probed on the intermediate level, no participant seemed to notice this pattern in their own behavior or incorporate it into a conscious strategy at the second level. 

Considering the color of the glass, the pattern changes.
\revisionEnhancement{
On the cognitive level, the connection between the color of the glass and its breakability is harder to identify than the relationship between weight and breakability. There’s no evidence of motor behavior effects related to color, nor is there a conscious strategy that links perceived color to motor control. To put it bluntly, the basic levels "agree" on weight but "disagree" on color, while the intermediate level lacks understanding of how to translate conscious knowledge into effective motor actions.
}

\revisionEnhancement{This dissociation pattern may be linked to prior knowledge and beliefs. Research shows that humans are skilled at aligning structural knowledge with data \citep{waldmann1996knowledge,waldmann2001estimating,lagnado2007beyond}. The weight-breakability link appears more plausible than the color-breakability connection, as greater weight may suggest thicker materials that break less easily. This difference in plausibility was a key factor in the experiment's design, which aimed to minimize interference from prior knowledge. Additionally, the uncertainty surrounding the color-breakability link prevents its effective integration into the sensorimotor optimization loop without further empirical data.}
 
\subsection{\revisionEnhancement{Role of Prior Information}}
\label{sect:role_prior_information}
\revisionEnhancement{
As expected, our results suggest a possible prior expectation about a link between glass weight and breakability; people may have experienced heavier glasses as being sturdier. Examining the color-breakability relation, the discovery-based sensorimotor causal representation analysis did not identify any relation between the two variables. This can be seen as evidence against a relevant prior for this relationship. Since the learning of the "novel" color-breakability relation was of special importance for us, we performed pre-tests to determine the degree of influence of color on the breakability in a way that the causal learning process should be clearly observable. In particular, we wanted to avoid that many of our subjects recognize the role of color already in the first session (raw). With hindsight, we may have been too cautious, as half of our subjects were unable to recognize the influence of color throughout the entire experiment. Furthermore, at the sensorimotor level, color is even a greater problem, since it not only has to "recognize" that color has an influence but also to find an at least approximately correct quantitative mapping. However, since the sensorimotor level seems to recognize, in principle, that there is some influence of color (see \cref{sect:sensorimotor_level}), we would expect that setting the gain of color higher in our experimental setup would enable both the cognitive level and the sensorimotor level to learn this causal relation.\\
At last, there seems to be an erroneous prior for the weight-color relation. This might stem from the fact that color is sometimes used as an indicator of continuous variables, such as weight. However, as there are also examples against this relation, it seems more plausible that subjects just assumed that the most apparent features are linked (see end of \cref{sect:conscious_cognitive_level}).

Due to the experimental design, it was not possible to gather information about the expectations of each participant. This also implies that we cannot infer how much learning happened during the first session. Interrogating participants at the start of the experiment would introduce a bias and alter the level of naivety that participants exhibit. Of course, this raises the question of whether learning success can be measured as a difference from zero or if a different level is more appropriate. Ultimately, we cannot answer this question without gathering more data about prior expectations.
}
\subsection{\revisionEnhancement{Comparison With Machine Algorithms}}
\label{sect:comparison_with_machine_algorithms}
\revisionEnhancement{

The results from the idealized set of variables highlight two key points: First, the experimental data are sufficient for complete learning, despite the limitations of a small sample size of 3 × 80 trials. Second, the learning rate of algorithms is comparable to that of human subjects, although algorithms typically outperform humans in identification. However, when both are provided the same variable set, the algorithms’ performance declines with low sample volumes, making them less stable in recovering causal relations, especially in identifying weak relationships, and ultimately end at a similar identification rate to humans (weight-breakability) or even worse (color-breakability). The recovered direct link between force\_value and the outcome of the experiment is important, as this indicates that the influence from the produced force on the state of the glass (broken/healthy) is recoverable. This indicates that our experimental paradigm works (see problems~\ref{item:sensorimotor_causal_representation},\ref{item:sensorimotor_relevance},\ref{item:observable_sensorimotor_representation}).

Next, we consider the causal parents of force\_value in the graph. Even if the human sensorimotor representation fails to grasp the color-force relation (which is often not learned by participants), the machine algorithm may detect a causal link between color and glass\_ok. This could suggest an ideal learner might view the sensorimotor representation as imperfect. However, since algorithms are applied across all sessions for each subject, potential learning effects must be taken into account. Thus, the factors affecting force\_value likely vary qualitatively and quantitatively within the dataset, leading us to avoid interpreting these relations based on the results.

}

\subsection{\revisionEnhancement{Conclusion}}
\label{sect:conclusion}

 \revisionEnhancement{

In this paper, we introduce a novel experimental paradigm to investigate human causal structure learning at both the cognitive and sensorimotor levels. This is the first study to explore causal learning on the sensorimotor level using a naturalistic task, which necessitates learning the same causal structure for successful task completion. Our exploratory study reveals that: a) causal structure learning is achievable in a sensorimotor context, b) it occurs at both cognitive and sensorimotor levels, and c) while the cognitive level may influence the sensorimotor level, both can operate largely independently in processing causal structures.

Several aspects deserve a closer investigation in future research: 
First, why is there a problem with the unique identification of causal directions? 
 Although our subjects could have determined the correct causal direction through commonsense reasoning about the task and experimental setup, as well as by utilizing the specific structure of the common-effect causal structure, they insisted until the very end of the experiment that they considered both causal directions equally possible.
 Second, it would be interesting to explore how far sensorimotor causal structure learning is possible with respect to other object properties, such as causal relations between shape and elasticity.
 Third, it would be desirable to investigate the influence of a substantially extended training period on causal learning.
 Not only because it is known that sensorimotor learning in general can profit from longer training than in the current experiment.
 Longer training would also provide us with more data for applying causal discovery to the sensorimotor data.
 Finally, a further obvious extension in future research of sensorimotor causal structure learning is the design of tasks and experimental setups that allow for and require interventions in the Pearlian sense for the unique identification of the causal structure. This also includes conducting an experiment with more participants to validate the findings, particularly given the small effect sizes suggested by our analysis (e.g., conditional mutual information across sessions).
 
In summary, the present results reinforce our conviction that the sensorimotor domain offers a valuable testbed for examining how causal information affects our perception of the environment and our motor interactions with it. 

}

\section{Conflict of interest statement}
The authors declare that the research was conducted in the absence of any commercial or financial relationships that could be construed as a potential conflict of interest.

\section{Author contributions}
NB: Conceptualization, Data curation, Formal analysis, Investigation, Methodology, Software, Visualization, Writing - original draft, Validation;
CZ: Conceptualization, Funding acquisition, Methodology, Writing - original draft, Writing - review \& editing;
JM: Conceptualization (VR setup), Software, Writing - original draft;
KS: Funding acquisition, Project administration, Supervision, Writing - review \& editing, Resources;

\section{Funding}
This work has been supported by the German Research Foundation DFG, as part of Collaborative Research Center (Sonderforschungsbereich) 1320 Project-ID 329551904 "EASE - Everyday Activity Science and Engineering", University of Bremen (\href{http://www.ease-crc.org/}{http://www.ease-crc.org/}). The research was conducted in subproject H01 "Sensorimotor and Causal Human Activity Models for Cognitive Architectures".

\section{Acknowledgments}
Usage and optimization of the causal discovery algorithms were possible thanks to discussion and support of Konrad Gadzicki (EASE - subproject H03 "Discriminative and Generative Human Activity Models for Cognitive Architectures")

\section{Supplemental Data}
\revisionEnhancement{Comprehensive information considering the experimental trial design and the questionnaire employed is available in the supplementary file. Additionally, numerical details concerning the causal variables/relationships, further causal discovery results, as well as used parameters for the phantom and algorithms are provided therein.}

\section{Ethics}
The experiments have been approved by the ethics commitee of the university of Bremen regarding the project H01 "Sensorimotor and Causal Human Activity Models for Cognitive Architectures" as part of Collaborative Research Center (Sonderforschungsbereich) "EASE - Everyday Activiy Science and Engineering".

\clearpage

\bibliographystyle{unsrt}
\bibliography{references}

\clearpage

\appendix

\section{On the Bayesian Causal Inference model of human perception}\label{sec:bci_discussion}
There is a fundamental difference between the structural discovery approach suggested here and the Bayesian Causal Inference (Bayesian CI) model of multisensory perception and sensorimotor adaptation~\citep{koerding2007causalMultisensory,shams2022bayesiancausalInference}.
In a strict sense, Bayesian CI is not a causal discovery method. 
It works on a single trial whereas our approach makes use of the statistics of a set of data (from multiple trials of the experiment).
Furthermore, in Bayesian CI there are two prespecified hypotheses for the causal structure which have been \emph{previously} learned and are then used by BCI. \cite{shams2022bayesiancausalInference} state, it is assumed that "The competing causal structures [...] have been learned by experience or over the course of evolution". Therefore, Bayesian CI describes, how the system makes use of the competing causal structures once they have been established (e.g. by some form of causal structure learning). Whereas in causal discovery there are multiple causal structures (the set of all causal structures that are possible with the given set of variables) and by learning from data it has to be determined which of them is the correct one.
In our approach, the underlying causal structure is not assumed to vary from trial to trial but is assumed to be (approximately) fixed throughout a period of the experiment.
Its identification requires a sufficiently large set of data and cannot be obtained from the data of a single trial.
Our method is a classical causal discovery approach in the sense that we have available a set of covariating values of causal variables (data from the trials of one or more sessions of our experiment) and try to identify the underlying causal structure.


\section{Experiment trial design}\label{appendix:experiment_trial_design}

The VR setting is made up of two plates, and indicator bar, and a closed hand holding a glass of wine as depicted in \cref{fig:vr_env}. A trial consists of placing the glass on the right plate, moving it on the left plate and touch the bar at the top at some point in between. 

The subject is instructed not to break the glass and to perform the states \textit{Transport Pre} and \textit{Transport Post} (see \cref{enum:methods_transport-pre} and \cref{enum:methods_transport-post}) as fast as possible. The glass will break if the variable \textit{result} in the causal model becomes zero, i.e. the haptic device registers a force above a given threshold (see \textbf{\nameref{appendix:experiment-causal_relations}}). 

To guide the subject, the next target to touch (plate/bar) appears blue until touched. Once it has touched by the glass, the start (right) plate will first become yellow then green, to indicate the start of the trial. The end (left) plate will act equivalent before the subject raises the glass to end the trial. The whole trial can be seen as the sequence of the following states:
\begin{enumerate}
    \item \textbf{Inter Trial} The glass is weightless. The subject can move the glass freely. The start plate becomes blue.
    \item \textbf{Touching Start Plate} The glass is placed on top of the start plate, which in turn first becomes yellow, then green. The bar at the top becomes blue once the glass does not touch the plate anymore.
    \item \label{enum:methods_transport-pre}\textbf{Transport Pre} The glass has to touch the bar at the top, which turns yellow once touched.
    \item \label{enum:methods_transport-post}\textbf{Transport Post} The glass has to touch the end plate.
    \item \textbf{Touching End Plate} The glass is placed on top of the end plate, which in turn first becomes yellow, then green. Afterwards the subject has to stop touching the plate. The glass breaks if the subject applies too much force during this state.
    \item \textbf{End} The glass becomes weightless and the subject gets feedback about the used force and whether the trial was successful or not.
\end{enumerate}

\clearpage

\section{Causal relations}\label{appendix:experiment-causal_relations}

The causal relations that govern the relevant features in the simulation are given by \crefrange{eq:causal_sampled_features}{eq:causal_color_offset} (note that the prefix \textit{glass} has been omitted for readability). See \cref{figure:ground_truth} to for a better overview of the causal information flow between the variables.

The exogenous features are sampled as

\begin{equation}
    \label{eq:causal_sampled_features}
    \begin{array}{lcl}
    \mathit{force} & ~ & \text{rendered force by \textit{PHANToM} in Newton} \\
    \mathit{color} & \sim & \mathcal{U}(\set{\text{red, green, blue}}) \\
    \mathit{weight} & \sim & \mathcal{U}([0.2616, 1])
\end{array}
\end{equation}

The weight is rendered by the haptic device as a force vector pointing directly downwards and is therefore able to simulate the weight of the object. The actual force acting on the subject in Newton is given by $\mathit{weight} \cdot 3N$. \textit{force\_threshold} and subsequently \textit{result} are calculated as
\begin{equation}
    \label{eq:causal_force_threshold}
    f_{\mathit{force\_threshold}}(\mathit{color}, \mathit{weight}) = (2.5\cdot \mathit{weight} + \mathit{color\_offset}(\mathit{color}))\cdot 0.8241
\end{equation}
and
\begin{equation}
    \label{eq:causal_result}
    f_\mathit{result}(\mathit{force\_threshold}, \mathit{force}) = 
    \begin{cases}
         1, & \text{for } \mathit{force} \leq \mathit{force\_threshold}\\
         0, & \text{else} ,
    \end{cases}
\end{equation}
with 
\begin{equation}
    \label{eq:causal_color_offset}
    \mathit{color\_offset}(\mathit{color}) = 
    \begin{cases}
        -0.254 & \text{for } \mathit{color} = \text{red}\\
        -0.069 & \text{for } \mathit{color} = \text{green}\\
        0.116 & \text{for } \mathit{color} = \text{blue}\\ .
    \end{cases}
\end{equation}

If $f_\mathit{result}(\mathit{force\_threshold}, \mathit{force}) = 1$, the glass remains intact. Otherwise, it breaks. One can imagine the relation as
\begin{equation}
    \text{Glass breaks} = 1 - f_\mathit{result}(\mathit{force\_threshold}, \mathit{force}).
\end{equation}

The color-specific offsets for \textit{force\_threshold} and the functional form were empirically determined to produce a noticeable difference in the breakability of the glass for each color without being too obvious.
After we ran the study with two test participants, we had to correct the threshold in \cref{eq:causal_force_threshold} by a factor of 0.8241 because the glass was too sturdy. We verified the design with one last test participant.

\clearpage

\section{Causal Discovery Results}\label{appendix:causal_discovery_results}
Whenever we applied causal discovery to data, we used the FCI, PC and FGES algorithms. All results we report are similar based on the different algorithms using the specified criteria as can be seen in the following figures.
\begin{figure}[h]
    \centering
    \begin{subfigure}[b]{0.25\textwidth}
        \centering
        \includegraphics[width=\linewidth]{participantGraph-session_type_raw_fci-alpha_05.png}
    \end{subfigure}
    \hfill
    \begin{subfigure}[b]{0.25\textwidth}
        \centering
        \includegraphics[width=\linewidth]{participantGraph-session_type_train_fci-alpha_05.png}
    \end{subfigure}
    \hfill
    \begin{subfigure}[b]{0.25\textwidth}
        \centering
        \includegraphics[width=\linewidth]{participantGraph-session_type_test_fci-alpha_05.png}
    \end{subfigure}
    \centering
    \begin{subfigure}[b]{0.25\textwidth}
        \centering
        \includegraphics[width=\linewidth]{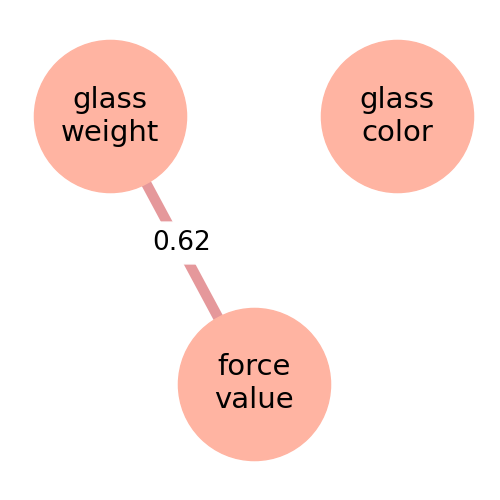}
    \end{subfigure}
    \hfill
    \begin{subfigure}[b]{0.25\textwidth}
        \centering
        \includegraphics[width=\linewidth]{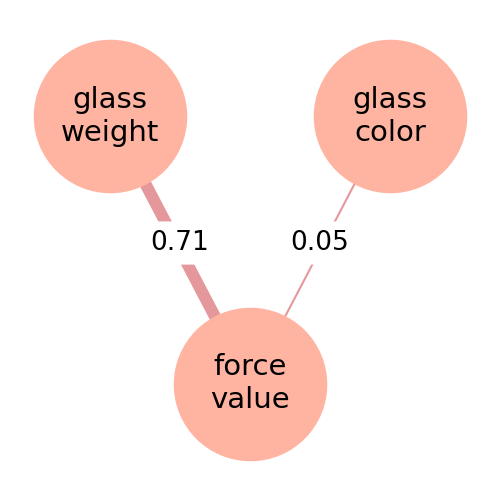}
    \end{subfigure}
    \hfill
    \begin{subfigure}[b]{0.25\textwidth}
        \centering
        \includegraphics[width=\linewidth]{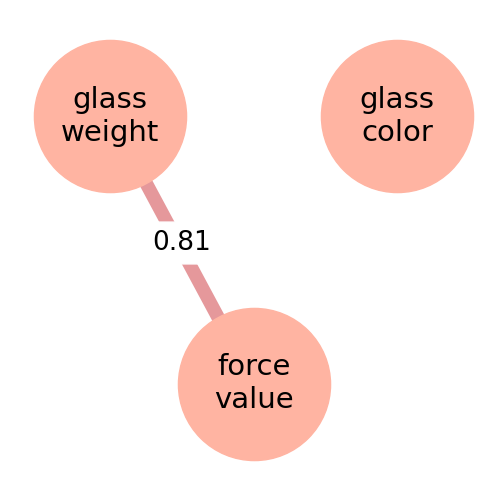}
    \end{subfigure}

\centering
    \begin{subfigure}[b]{0.25\textwidth}
        \centering
        \includegraphics[width=\linewidth]{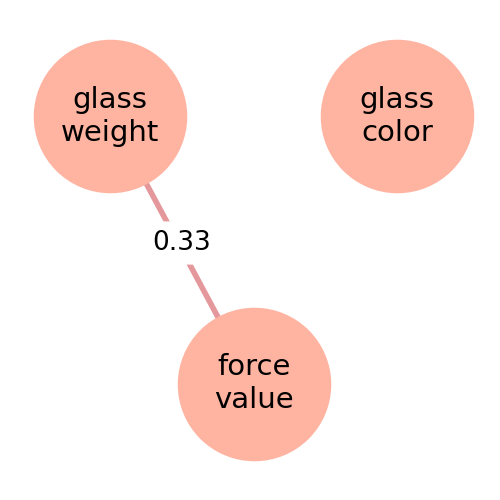}
    \end{subfigure}
    \hfill
    \begin{subfigure}[b]{0.25\textwidth}
        \centering
        \includegraphics[width=\linewidth]{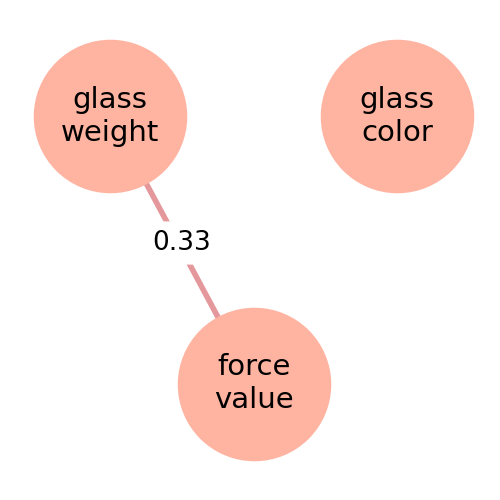}
    \end{subfigure}
    \hfill
    \begin{subfigure}[b]{0.25\textwidth}
        \centering
        \includegraphics[width=\linewidth]{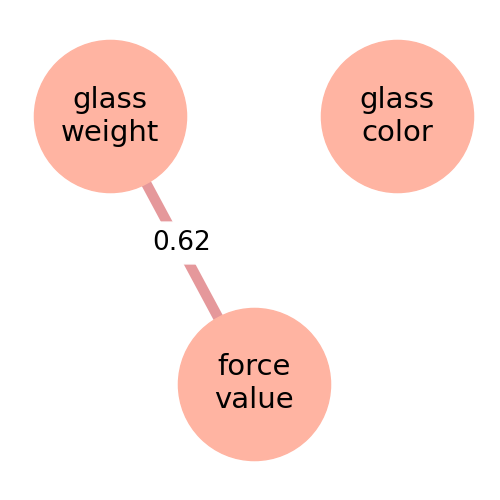}
    \end{subfigure}

    \centering
    \caption{Sensorimotor causal graphs identified by the causal discovery algorithm (sessions from left to right: raw, train, test, algorithms from top to bottom: FCI, PC, FGES)}
    \label{appendix:causal_discovery_sensorimotor}
\end{figure}

\begin{figure}[h]
    \centering
    \begin{subfigure}[b]{0.3\textwidth}
        \centering
        \includegraphics[width=\linewidth]{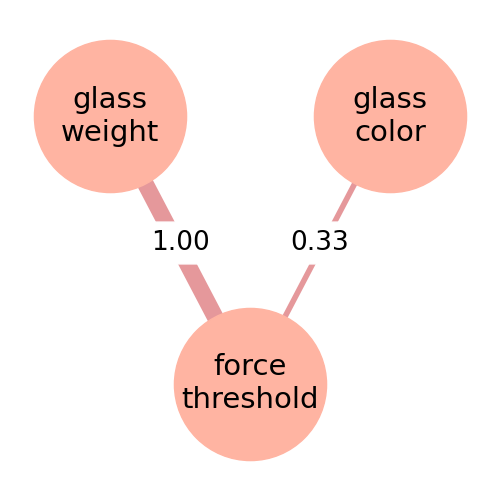}
    \end{subfigure}
    \hfill
    \begin{subfigure}[b]{0.3\textwidth}
        \centering
        \includegraphics[width=\linewidth]{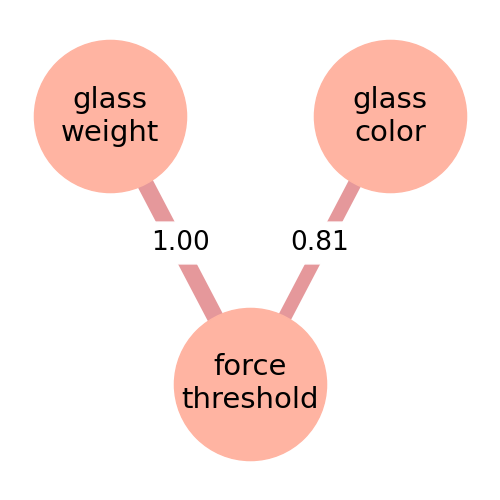}
    \end{subfigure}
    \hfill
    \begin{subfigure}[b]{0.3\textwidth}
        \centering
        \includegraphics[width=\linewidth]{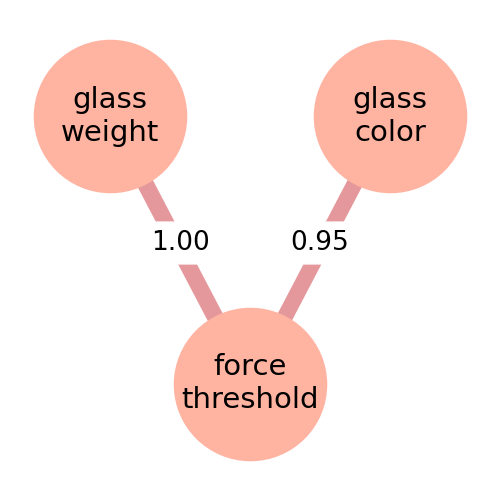}
    \end{subfigure}

    \centering
    \begin{subfigure}[b]{\textwidth}
        \centering
        \includegraphics[width=\linewidth]{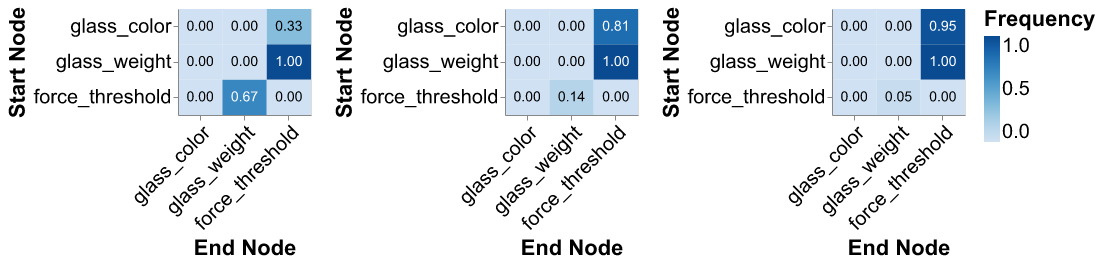}
    \end{subfigure}
    
    \caption{Causal machine learning with discovery algorithms across the three sessions with an idealized variable set for FCI algorithm. Upper part: Causal connectivity. Lower part: Causal directions.}
    \label{appendix:causal_discovery_fci_idealized}
\end{figure}

\begin{figure}[h]
    \centering
    \begin{subfigure}[b]{0.3\textwidth}
        \centering
        \includegraphics[width=\linewidth]{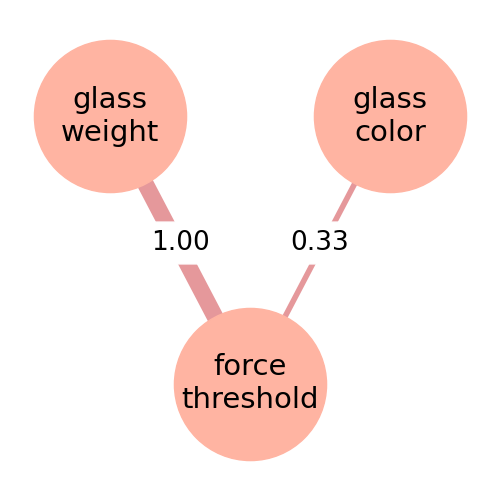}
    \end{subfigure}
    \hfill
    \begin{subfigure}[b]{0.3\textwidth}
        \centering
        \includegraphics[width=\linewidth]{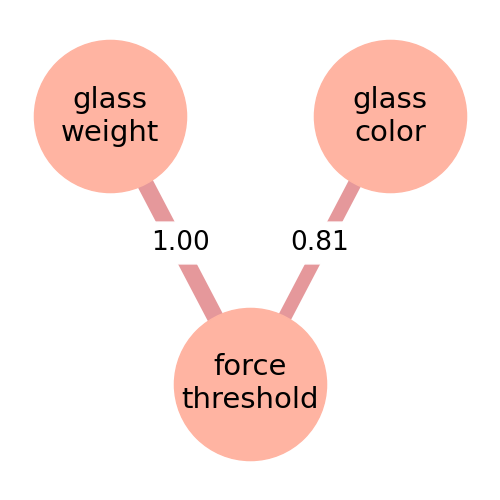}
    \end{subfigure}
    \hfill
    \begin{subfigure}[b]{0.3\textwidth}
        \centering
        \includegraphics[width=\linewidth]{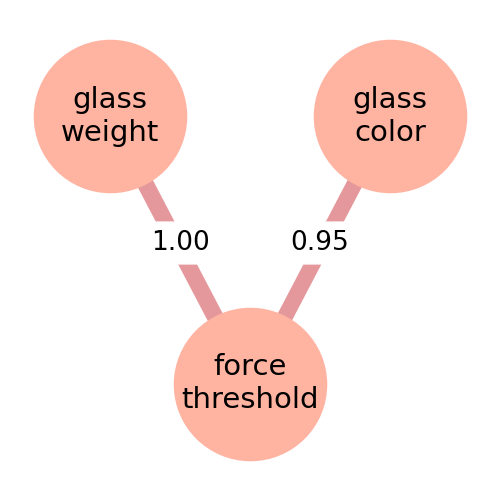}
    \end{subfigure}

    \centering
    \begin{subfigure}[b]{\textwidth}
        \centering
        \includegraphics[width=\linewidth]{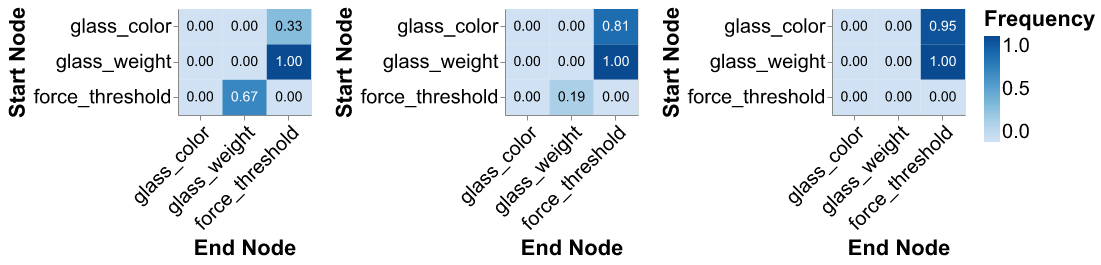}
    \end{subfigure}
    
    \caption{Causal machine learning with discovery algorithms across the three sessions with an idealized variable set for PC algorithm. Upper part: Causal connectivity. Lower part: Causal directions.}
    \label{appendix:causal_discovery_pc_idealized}
\end{figure}

\begin{figure}[h]
    \centering
    \begin{subfigure}[b]{0.3\textwidth}
        \centering
        \includegraphics[width=\linewidth]{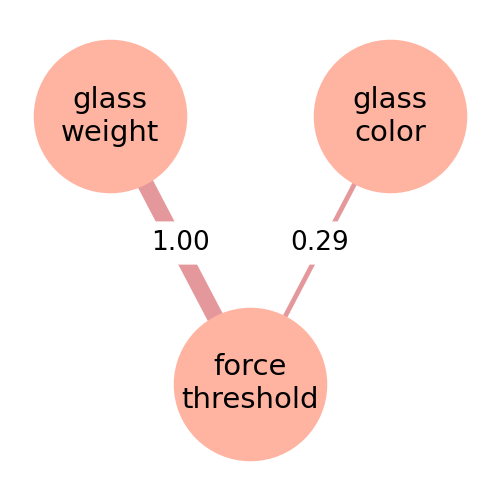}
    \end{subfigure}
    \hfill
    \begin{subfigure}[b]{0.3\textwidth}
        \centering
        \includegraphics[width=\linewidth]{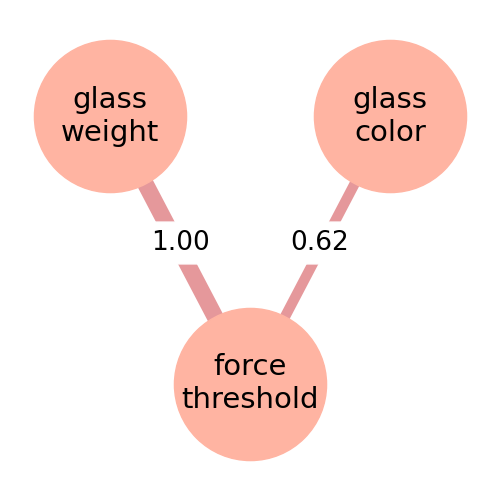}
    \end{subfigure}
    \hfill
    \begin{subfigure}[b]{0.3\textwidth}
        \centering
        \includegraphics[width=\linewidth]{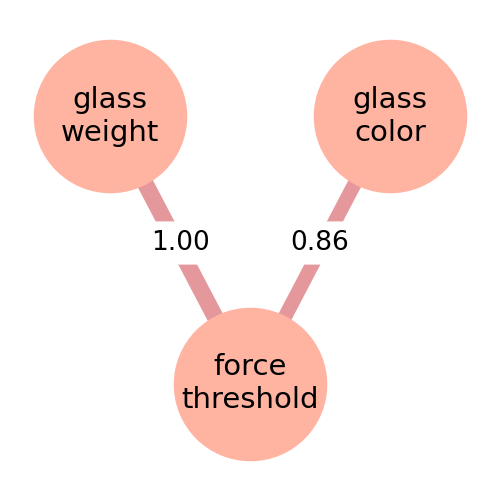}
    \end{subfigure}

    \centering
    \begin{subfigure}[b]{\textwidth}
        \centering
        \includegraphics[width=\linewidth]{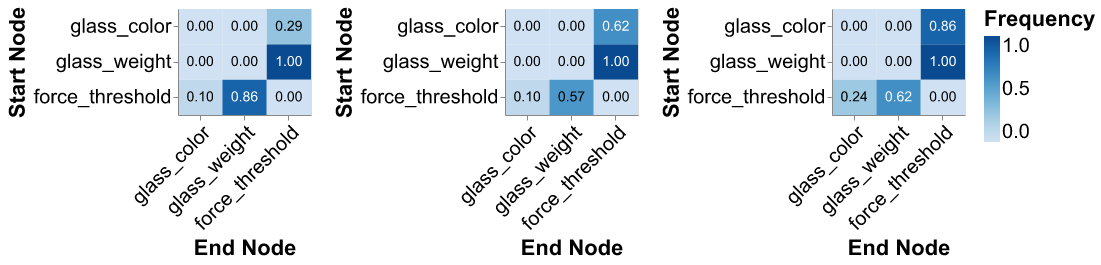}
    \end{subfigure}
    
    \caption{Causal machine learning with discovery algorithms across the three sessions with an idealized variable set for FGES algorithm. Upper part: Causal connectivity. Lower part: Causal directions.}
    \label{appendix:causal_discovery_fges_idealized}
\end{figure}

\begin{figure}[h]
    \centering
    \begin{subfigure}[b]{0.25\textwidth}
        \centering
        \includegraphics[width=\linewidth]{algorithmGraph-session_type_raw-fci.png}
    \end{subfigure}
    \hfill
    \begin{subfigure}[b]{0.25\textwidth}
        \centering
        \includegraphics[width=\linewidth]{algorithmGraph-session_type_raw_train-fci.png}
    \end{subfigure}
    \hfill
    \begin{subfigure}[b]{0.25\textwidth}
        \centering
        \includegraphics[width=\linewidth]{algorithmGraph-session_type_raw_train_test-fci.png}
    \end{subfigure}
    \centering
    \begin{subfigure}[b]{0.25\textwidth}
        \centering
        \includegraphics[width=\linewidth]{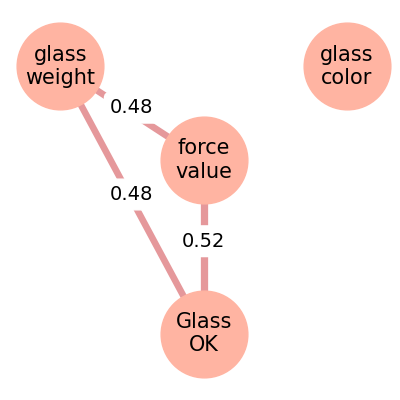}
    \end{subfigure}
    \hfill
    \begin{subfigure}[b]{0.25\textwidth}
        \centering
        \includegraphics[width=\linewidth]{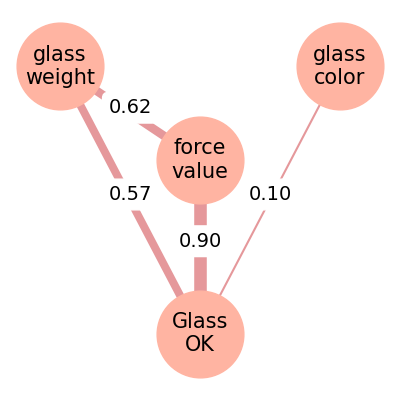}
    \end{subfigure}
    \hfill
    \begin{subfigure}[b]{0.25\textwidth}
        \centering
        \includegraphics[width=\linewidth]{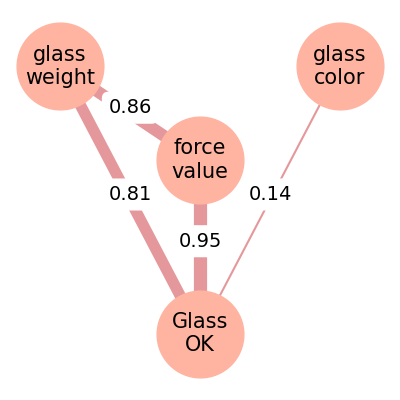}
    \end{subfigure}

    \centering
    \begin{subfigure}[b]{0.25\textwidth}
        \centering
        \includegraphics[width=\linewidth]{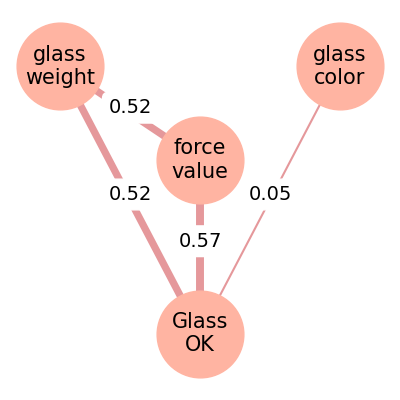}
    \end{subfigure}
    \hfill
    \begin{subfigure}[b]{0.25\textwidth}
        \centering
        \includegraphics[width=\linewidth]{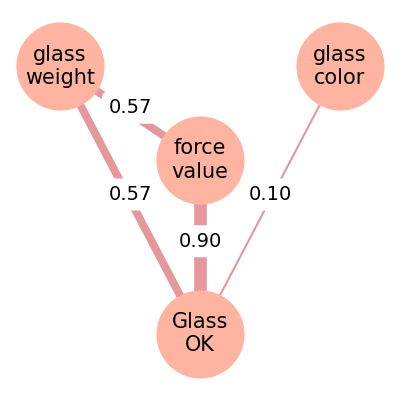}
    \end{subfigure}
    \hfill
    \begin{subfigure}[b]{0.25\textwidth}
        \centering
        \includegraphics[width=\linewidth]{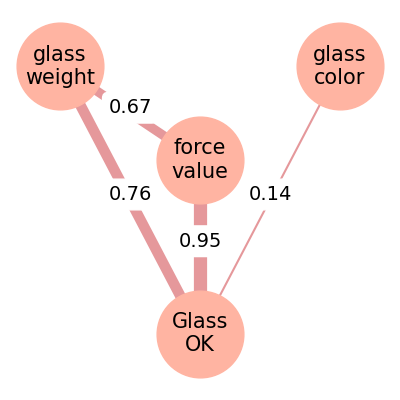}
    \end{subfigure}

    \centering
    \caption{Causal relations discovered by a machine algorithm from the observable data available to humans in the experiment (sessions from left to right: raw, train, test, algorithms from top to bottom: FCI, PC, FGES)}
    \label{appendix:causal_discovery_comparison_realistic}
\end{figure}

\FloatBarrier
\section{Parameters}

\begin{table}[h!]
    \centering
    
    \begin{tabular}{@{}ll@{}}
        \toprule
        \textbf{Parameter} & \textbf{Value} \\\midrule
        \rowcolor{highlightTable}
        Model type & Phantom Premium 1.5 \hfill\\
        Device version & 3.50.0 \\
        \rowcolor{highlightTable}
        Driver version & 5.1.7 \\
        Firmware version & 12 \\
        \midrule
        \rowcolor{highlightTable}
        Force ramping rate & 0.5 [N/s] \\
        Update rate & 1000 [Hz] \\
        \rowcolor{highlightTable}
        Stiffness & 1 \\
        Static friction & 0 \\
        \rowcolor{highlightTable}
        Dynamic friction & 0 \\
    \bottomrule
    \end{tabular}
    \caption{Specifications and parameter used for the Phantom haptic device}
    \label{tab:appendix_phantom_specifications}
    
\end{table}

\begin{table}[h]
    \centering
    
    \begin{tabular}{@{}lll@{}}
        \toprule
        \textbf{Algorithm} & \textbf{Hyperparameter} & \textbf{value} \\\midrule
        \rowcolor{highlightTable}	
        	& resamplingWithReplacement & yes\\
        	& addOriginalDataset & yes\\
        \rowcolor{highlightTable}	
        	& saveBootstrapGraphs & no\\
        	& percentResampleSize & 80\\
        \rowcolor{highlightTable}	
        	& numberResampling & 60\\
        	& resamplingEnemble & 1\\
        \rowcolor{highlightTable}
            & seed & -1\\
        	& timeLag & 0\\
        \rowcolor{highlightTable}
        	& stableFAS & yes\\
            & discretize & yes\\
        \rowcolor{highlightTable}	
            & numCategoriesToDiscretize & 3\\
        	& alpha & 0.01\\
        \rowcolor{highlightTable}	
        	& depth & -1\\
        	& maxPathLength & -1\\
        \rowcolor{highlightTable}	
        FCI	& completeRuleSetUsed & yes\\
        FCI	& doDiscriminatingPathRule & yes\\
        \rowcolor{highlightTable}	
        FCI	& possibleMsepDone & yes\\
        \rowcolor{highlightTable}	
        PC & meekPreventCycles& yes\\
        PC & useMaxPHeuristic& no\\
        \rowcolor{highlightTable}
        PC & conflictRule& 1\\
         \bottomrule
    \end{tabular}
    
    \caption{Used hyperparameter for the FCI and PC algorithm. The column \textit{Algorithm} denotes a value/hyperparameter, which is only set for specific criteria. If left empty, the setting is used for both algorithms.}
    \label{tab:appendix_hyperparams-pc_fci}
\end{table}

\begin{table}[h]
    \centering
    
    \begin{tabular}{@{}lll@{}}
        \toprule
        \textbf{Hyperparameter} & \textbf{value} \\\midrule
        \rowcolor{highlightTable}	
        \rowcolor{highlightTable}	
        resamplingWithReplacement& yes\\
        addOriginalDataset& yes\\
        \rowcolor{highlightTable}	
        saveBootstrapGraphs& no\\
        percentResampleSize& 80\\
        \rowcolor{highlightTable}	
        numberResampling& 60\\
        resamplingEnsemble& 1\\
        \rowcolor{highlightTable}
        seed& -1\\
        \rowcolor{highlightTable}	
        timeLag& 0\\
        faithfulnessAssumed& no\\
        meekVerbose& no\\
        \rowcolor{highlightTable}	
        parallelized& no\\
        precomputeCovariances& yes\\
        \rowcolor{highlightTable}	
         symmetricFirstStep& no\\
        \rowcolor{highlightTable}	
        structurePrior& 1.0\\
        maxDegree& 1000\\
        \rowcolor{highlightTable}	
        penaltyDiscount& 2.0\\
         \bottomrule
    \end{tabular}
    
    \caption{Used hyperparameter for the FGES algorithm}
    \label{tab:appendix_hyperparams-fges}
\end{table}
\FloatBarrier
\clearpage
\section{Questionnaire (german)}

\textbf{Fragen}~\\
\begin{itemize}
    \item Glas-Farbe beeinflusst
        \begin{AutoMultiColItemize}
            \renewcommand\labelitemi{\LARGE$\square$}
            \item[\LARGE$\square$] Glas-Gewicht
            \item[\LARGE$\square$] Glas-Zerbrechlichkeit
            \item[\LARGE$\square$] Nichts
        \end{AutoMultiColItemize}
    \item Glas-Gewicht beeinflusst
        \begin{AutoMultiColItemize}
            \item[\LARGE$\square$] Glas-Farbe
            \item[\LARGE$\square$] Glas-Zerbrechlichkeit
            \item[\LARGE$\square$] Nichts
        \end{AutoMultiColItemize}
    \item Glas-Zerbrechlichkeit beeinflusst
        \begin{AutoMultiColItemize}
            \item[\LARGE$\square$] Glas-Gewicht
            \item[\LARGE$\square$] Glas-Farbe
            \item[\LARGE$\square$] Nichts
        \end{AutoMultiColItemize}
\end{itemize}

\vspace{3cm}

\begin{center}
    \includegraphics[width=\textwidth/3]{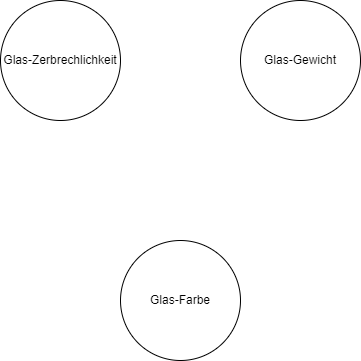}
\end{center}

\end{document}